\documentclass[twocolumn,english,aps,reprint, superscriptaddress,showpacs,showkeys]{revtex4-2}
\usepackage{amsmath,amssymb,bbm,mathrsfs,bm,braket,color,graphicx,comment}
\usepackage[colorlinks,citecolor=blue,urlcolor=blue,linkcolor = blue]{hyperref}
\usepackage{soul}

\usepackage[mathscr]{euscript}

\usepackage{hyperref}

\usepackage{tikz}

\usepackage{xcolor}
\usepackage{multirow}

\newcommand{\squaredots}{
    \vspace{-.175em}
    \tikz[line cap=round, line join=round]{
    \draw[black] (0ex,0ex) -- (0ex,0.8ex) --  (0.8ex,0.8ex) --  (0.8ex,0ex) -- cycle;
    \draw[color=black, fill=black] (0ex,0ex) circle (0.175ex);
    \draw[color=black, fill=black] (0ex,0.8ex) circle (0.175ex);
    \draw[color=black, fill=black] (0.8ex,0ex) circle (0.175ex);
    \draw[color=black, fill=black] (0.8ex,0.8ex) circle (0.175ex);
    }
}

\begin{document}

\title{Rydberg blockade based parity quantum optimization}

\author{Martin Lanthaler}
\thanks{These authors contributed equally to this work. Corresponding author: martin.lanthaler@uibk.ac.at}
\affiliation{Institute for Theoretical Physics, University of Innsbruck, A-6020 Innsbruck, Austria}
\author{Clemens Dlaska}
\thanks{These authors contributed equally to this work. Corresponding author: martin.lanthaler@uibk.ac.at}
\affiliation{Institute for Theoretical Physics, University of Innsbruck, A-6020 Innsbruck, Austria}
\author{Kilian Ender}
\affiliation{Institute for Theoretical Physics, University of Innsbruck, A-6020 Innsbruck, Austria}
\affiliation{Parity Quantum Computing GmbH, A-6020 Innsbruck, Austria}
\author{Wolfgang Lechner}
\affiliation{Institute for Theoretical Physics, University of Innsbruck, A-6020 Innsbruck, Austria}
\affiliation{Parity Quantum Computing GmbH, A-6020 Innsbruck, Austria}

\begin{abstract}
    We present a scalable architecture for solving higher-order constrained binary optimization problems on current neutral-atom hardware operating in the Rydberg blockade regime. In particular, we formulate the recently developed parity encoding of arbitrary connected higher-order optimization problems as a maximum-weight independent set (\textsf{MWIS}) problem on disk graphs, that are directly encodable on such devices. Our architecture builds from small \textsf{MWIS} modules in a problem-independent way, crucial for practical scalability. 
\end{abstract}

\maketitle
\textit{Introduction.---} With the recent advances in experiments, programmable arrays of Rydberg atoms have become a versatile platform for quantum computing and quantum simulation~\cite{Saffman2010, Henriet2020, Scholl2021, Bluvstein2021, Ebadi2021}. In particular, the platform has been identified as a prime candidate for tackling combinatorial optimization problems using quantum annealing and variational quantum algorithms~\cite{Ebadi2022, Graham2022}. Many combinatorial optimization problems are known to be hard to solve for classical computers and thus current quantum computing efforts are directed toward exploring potential quantum speedups. \par
A key challenge to reach this goal, especially in the current era of noisy intermediate scale quantum (NISQ) devices~\cite{Preskill2018_nisq}, is to make efficient use of the available physical resources. To this end, it is of crucial importance to match the strengths of a particular physical platform to the computational problem under consideration.  In the realm of quantum optimization, it was recently shown that the Rydberg platform provides a natural, overhead-free match for encoding maximum-weight independent set (\textsf{MWIS}) problems on disk graphs~\cite{Pichler2018,Ebadi2022} based on the so-called Rydberg blockade mechanism~\cite{Jacksch2000, Lukin2001,Gaetan2009}. Even though solving this ``Rydberg-encoded'' \textsf{MWIS} problem has been shown to be NP-hard~\cite{Pichler2018_complexity}, it remains challenging to extend this approach restricted to disk graphs for encoding arbitrary optimization problems. This is especially true for problems that go beyond quadratic binary optimization (QUBO), i.e.,  allowing for higher-order interaction terms (hypergraphs) and side conditions (hard constraints). \par 
Here, we propose a scheme to overcome these limitations by utilizing the recently introduced parity encoding of arbitrary combinatorial optimization problems~\cite{Lechner2015, Ender2021}. Our scheme allows one to construct the problem-defining spin model from small, problem-independent \textsf{MWIS} blocks on a two-dimensional lattice geometry, paving the way for straightforward scalability. Furthermore, we equip our approach with a local compensation routine minimizing unwanted Rydberg-interaction-induced crosstalk and numerically demonstrate, that our procedure faithfully encodes the solution to the optimization problem as a \textsf{MWIS} ground state. \par


\begin{figure}[t]
\begin{centering}
\includegraphics[width = \columnwidth]{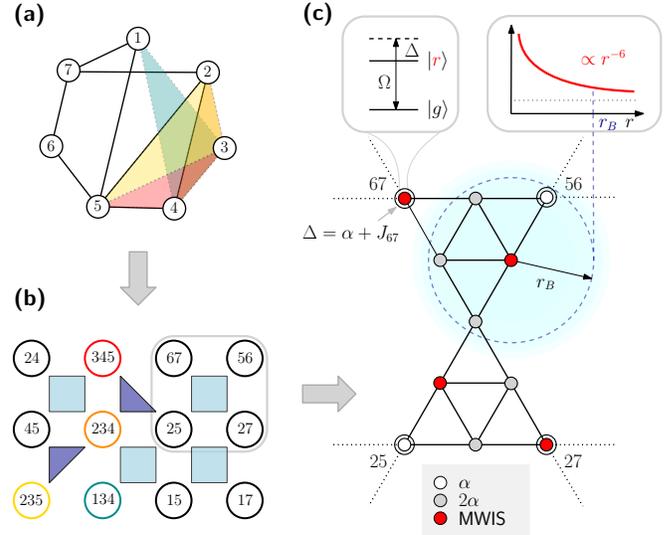}
\par\end{centering}
\protect\caption{\textit{Rydberg parity \textsf{MWIS} protocol.} Arbitrary combinatorial optimization problems (a) can be parity-transformed to a spin model utilizing only local-fields and quasi-local interactions (indicated by blue squares and triangles) (b). The resulting plaquette logic can be realized as a \textsf{MWIS} problem on disk graphs, readily implementable in state-of-the-art neutral atom devices (c). There, each vertex is represented by an atom, with states $\ket{0}$ and $\ket{1}$ encoded in electronic ground $\ket{g}$ and strongly interacting Rydberg states $\ket{r}$, coherently driven by Rabi frequency $\Omega$ and detunings $\Delta$. Strong van der Waals interactions energetically unfavour configurations where atoms in state $\ket{r}$ are closer than $r_B$.}
\label{fig:Fig1}
\end{figure}

\textit{Maximum-weight independent set with Rydberg atom arrays.---} A paradigmatic NP-complete problem is the so-called maximal independent set (\textsf{MIS}) problem~\cite{Karp1972}. Given a graph $G = (V, E)$, an independent set $S\subseteq V$ is a subset of vertices where no pair is connected via an edge $e\in E$. The problem of finding the independent set with maximal cardinality is called \textsf{MIS} problem. Assigning a weight $\Delta_v>0$ to each vertex generalizes the \textsf{MIS} problem to the \textsf{MWIS} problem that asks for the independent set with maximal total weight $W=\sum_{v\in S}\Delta_v$. The \textsf{MWIS} problem can be formulated as a spin model

\begin{equation}
    H_{\mathsf{MWIS}} = -\sum_{v\in V} \Delta_v \hat n_v +  \sum_{(v,w)\in E}U_{vw}\hat n_v \hat n_w,
    \label{eq:MWISHamiltonian}
\end{equation}
where each vertex is associated to a spin-1/2 system (with states $\ket{0}$ and $\ket{1}$) and $\hat n_v={\ket{1}_v}{\bra{1}}$. For $U_{vw}>\Delta_v>0$, $H_{\mathsf{MWIS}}$ energetically favors spin configurations to be in the state $\ket{1}$ and penalizes adjacent spins whenever they are in state $\ket{1}$.  Hence, the ground state of $H_\textsf{MWIS}$ is the sought-after \textsf{MWIS}. For disc graphs, these problems have been shown to be native to programmable arrays of neutral atoms~\cite{Pichler2018, Ebadi2022}, where individual atoms are trapped in optical tweezers and deterministically placed in two dimensions. Each atom realizes a qubit where the electronic ground state represents the state $\ket{0}$ and a highly excited Rydberg state represents the state $\ket{1}$. These states can be coherently coupled via laser light and induce strong dipolar interactions between pairs of atoms in state $\ket{1}$.

The corresponding Hamiltonian describing this system is given by  
\begin{equation}
    H_{\mathrm{Ryd}} =  \sum_i \left(\frac{\Omega_i}{2} \hat\sigma_x^{(i)} - \Delta_i \hat n_i\right) + \sum_{i<j}V(|\boldsymbol{x}_i-\boldsymbol{x}_j|)\hat n_i \hat n_j,
    \label{eq:RydbergHamiltonian}
\end{equation}
where $\Omega_i$ and $\Delta_i$ denote the Rabi frequency and laser detuning of atoms at position $\boldsymbol{x}_i$, respectively. Coherent laser coupling induces flips among the computational states, i.e., $\hat\sigma_x={\ket{0}}{\bra{1}}+{\ket{1}}{\bra{0}}$, and the interaction is assumed to be of isotropic van der Waals (vdW) form $V(|\boldsymbol{x}_i-\boldsymbol{x}_j|)=C_6/|\boldsymbol{x}_i-\boldsymbol{x}_j|^6$. The vdW interaction induces a strong energy penalty on nearby atoms in the Rydberg state. This gives rise to the so-called \textit{Rydberg blockade}~\cite{Jacksch2000,Lukin2001,Saffman2010} which prevents Rydberg excitations around a Rydberg-excited atom positioned at $\boldsymbol{x}_v$ within a characteristic distance called blockade radius $r_B(\Delta_v) = \left(C_6/\Delta_v\right)^{1/6}$. The blockade mechanism directly corresponds to disk graphs~\cite{Fishkin2004}, i.e., intersection graphs of a set of disks in the Euclidian plane, where the $v$th disk radius can be identified as the blockade radius $r_B(\Delta_v)$. Neglecting the long-range tails of the dipolar interaction, $H_\text{Ryd}(\Omega_i=0, \Delta_i>0)$ is able to directly realize Eq.~\eqref{eq:MWISHamiltonian} for disk graphs [see Fig.~\ref{fig:Fig1}(c)]. In order to solve the optimization problem one can for example devise a quantum annealing algorithm~\cite{Albash2018} that adiabatically connects the ground state of $H_\text{Ryd}(\Omega_i=0, \Delta_i<0)$, where all atoms are in state $\ket{0}$, to the solution-encoding ground state of $H_\text{Ryd}(\Omega_i=0, \Delta_i>0)$.\\
\begin{figure}[t]
    \centering
    \includegraphics[width = \columnwidth]{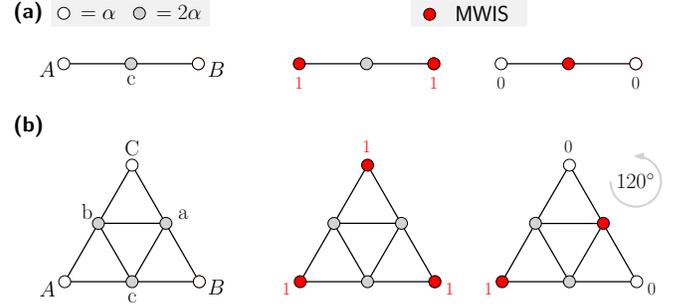}
    \caption{\textit{Basic parity \textsf{MWIS} building blocks}. (a) Two-body constraint $n_A\oplus n_B = 0$ as three-node MWIS problem with solutions. (b)  The three-body parity constraint $n_A\oplus n_B\oplus n_C = 1$ can be recast as a MWIS problem on six nodes. The four MWISs $\{A,B,C\}, \{A,a\}$, $\{B,b\}$ and $\{C,c\}$ are in one-to-one correspondence with the corresponding parity-constraint fulfilling assignments.}
    \label{fig:Fig2}
\end{figure}
\textit{Parity encoding as \textsf{MWIS}.---} As stated above, a remaining challenge is to utilize the Rydberg platform operating in the Rydberg blockade regime for tackling generic combinatorial optimization problems. These optimization problems can be cast in the form of an energy minimization of (classical) $N$-spin Hamiltonians
\begin{equation}\label{eq:problem_hamiltonian}
\begin{split}
    H_\text{problem} = & \sum_{i} J_{i} s_i +  \sum_{i<j} J_{ij} s_i s_j 
    \\& + \sum_{i<j<k} J_{ijk} s_i s_j s_k + \dots,
\end{split} 
\end{equation}
where ${s_i=\pm 1}$ denote spin variables and the coefficients $\lbrace J_i, J_{ij}, J_{ijk},\dots\rbrace$ describe local fields and arbitrarily long-ranged and higher-order interactions between spins [see Fig.~\ref{fig:Fig1}(a)]. Direct experimental implementations of Eq.~\eqref{eq:problem_hamiltonian}
with Rydberg atoms are, however, limited to rather specific optimization graphs due to the dependency on the hardware geometry and the polynomially decaying interaction strengths. The parity architecture \cite{Lechner2015,Ender2021} deals with this difficulty by encoding the relative orientation of original problem spins into parity-qubits, e.g.,
$ J_{ijk}\, s_i s_j s_k \mapsto J_m{\hat n_m} $, where each parity qubit is labeled by the corresponding problem spin indices ${m = (i,j,k)}$ with binary eigenvalues ${n_m = (1+{\prod_{i\in m}s_i})/2\in\{0,1\}}$. This transformation reduces multibody interactions to local fields which increases the number of qubits to the number $K\geq N$ of interactions present in the optimization problem. In order to stabilize the original $N$-spin code space, quasi-local three- or four-qubit constraints  on $2\times2$ plaquettes are introduced [see Fig.~\ref{fig:Fig1}(b)]. They are required to fulfill
\begin{equation}
\label{eq:BinaryConstraint}
 C_{\squaredots}:
\begin{cases}
n_A\oplus n_B \oplus n_C  = 1\\
n_A\oplus n_B \oplus n_C \oplus n_D = 0,
\end{cases}
\end{equation}
where $\oplus$ denotes addition modulo two.  Our goal now is to formulate the logic imposed by  $C_{\squaredots}$ as a \textsf{MWIS} spin-Hamiltonian  $H_{\squaredots}^{\textsf{MWIS}}$ [cf.~Eq.~\eqref{eq:MWISHamiltonian}] commensurate with the Rydberg platform operating in the blockade regime. The parity-encoded optimization problem then also corresponds to a \textsf{MWIS} problem providing the advantage of practical scalability in a problem-independent and modular way as
\begin{equation}\label{eq:localfield_hamiltonian}
    \hat H_\text{phys} = \sum_{m}^K J_{m}{\hat n_m} + \sum_{\squaredots}H_{\squaredots}^{\textsf{MWIS}},
\end{equation}
where the first sum contains the problem-defining local fields and the second sum runs over all 2 × 2 plaquettes, denoted by\squaredots.\par
\textit{Parity constraints as \textsf{MWIS}.---} We start from the simplest parity constraint $s_1s_2 = 1$ involving two parity qubits required to fulfill
\begin{equation}
n_A\oplus n_B = 0.
\label{eq:2bodyBinaryConstraint}
\end{equation}
This constraint can be formulated as a \textsf{MWIS} problem on a three-vertex path graph, where an auxiliary node $c$ is placed in between parity nodes $A$ and $B$, respectively [see Fig.~\ref{fig:Fig2}(a)]. Assigning weight $\alpha>0$ to the parity nodes while assigning weight $2\alpha$ to the auxiliary node gives rise to the \textsf{MWIS} cost function
\begin{equation}
    H_\text{2\textsf{MWIS}}= -\alpha(\hat n_A  + \hat n_B + 2\hat n_c ) + U( \hat n_{A} \hat n_{c} + \hat n_{B} \hat n_{c}),
\end{equation}
where $U>\alpha$. One can easily verify that this graph has two degenerate \textsf{MWIS}s [see Fig.~\ref{fig:Fig2}(a)], with the property that the parity-node assignments fulfill the desired constraint given by Eq.~\eqref{eq:2bodyBinaryConstraint}. \\
\begin{figure}[t]
\begin{centering}
\includegraphics[width =\columnwidth]{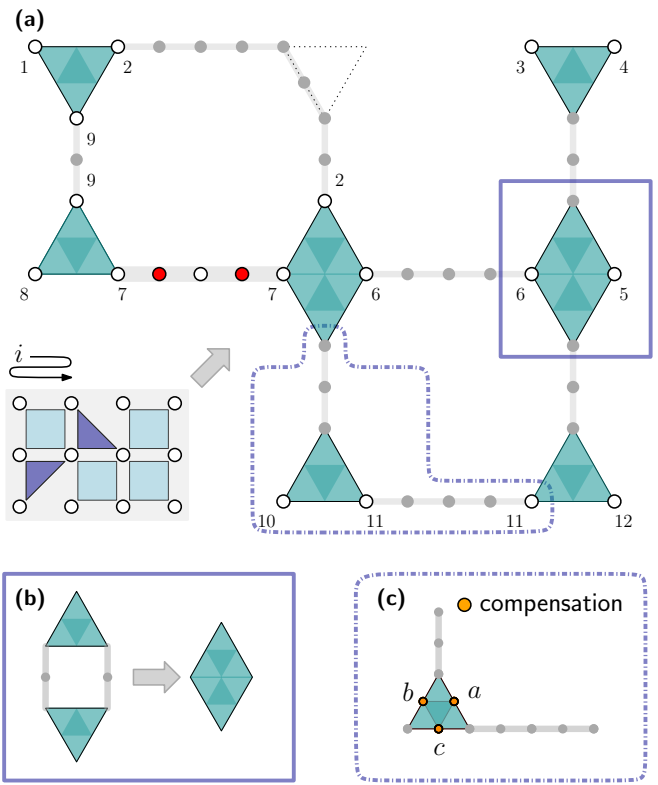}
\par\end{centering}
\protect\caption{\textit{Implementation.} (a) Example implementation for the parity layout shown in Fig.~\ref{fig:Fig1}. Modules are placed on a regular, two-dimensional grid and connected via links. Atoms at the module corners that are highlighted in white represent the parity qubits (cf.~bottom left panel).
Note, that the three-body constraint in the lower left corner is already implemented as part of the four-body module directly above.
(b) Whenever the construction requires two three-body modules stacked vertically, they can be simplified to a rhombic module requiring fewer atoms. 
(c) Module-link compensation in form of detunings is implemented in order to reduce unwanted crosstalk due to interaction tails. 
}
\label{fig:Fig3}
\end{figure}
The main building blocks relevant for the formulation of the parity architecture as
\textsf{MWIS} problem are the three- and four-body parity constraints of Eq.~\eqref{eq:BinaryConstraint}. Their logic can be constructed from three-body constraints where the four-body constraint is composed of two three-body constraints as
\begin{equation}
    (n_A\oplus n_B \oplus n_e = 1) \wedge
            (n_{e} \oplus n_{C} \oplus n_D = 1),
    \label{eq:4bodyBinaryConstraint}
\end{equation}
and $n_e$ denotes an auxiliary variable. This observation carries over to the \textsf{MWIS} representations of the constraints meaning that the \textsf{MWIS} formulation of the four-body constraint is the union of the \textsf{MWIS} graphs corresponding to the individual three-body constraints.\par
The three-body constraint $n_A\oplus n_B \oplus n_C = 1$ can be formulated as a \textsf{MWIS} problem on a six-vertex disk graph~\footnote{Note, that this is the minimum number of vertices which allow for such an encoding.} with three parity nodes $(A,B,C)$ and three auxiliary nodes $(a,b,c)$ as showcased in the leftmost panel of Fig.~\ref{fig:Fig2}(b). Parity nodes have weight $\alpha$, while auxiliary nodes have weight $2\alpha$, which gives rise to a \textsf{MWIS}  Hamiltonian [cf.~Eq.~\eqref{eq:MWISHamiltonian}] of the form
\begin{equation}
\label{eq:3MWIS}
    \begin{split}
    H_{3\mathsf{MWIS}} = &\,  \alpha (\hat n_A + \hat n_B + \hat n_C) \\ &+2\alpha (\hat n_a + \hat n_b + \hat n_c) \\ 
    & + U[\hat n_A(\hat n_b+\hat n_c) + \hat n_B(\hat n_a+\hat n_c)\\
    & + \hat n_C(\hat n_a+\hat n_b)+\hat n_a\hat n_b+\hat n_a\hat n_c+\hat n_b\hat n_c],
    \end{split}
\end{equation}
where it is straightforward to verify that all \textsf{MWIS}s have weight $3\alpha$ corresponding to the subsets $\{A,B,C\}$, $\{A,a\}$, $\{B,b\}$ and $\{C,c\}$ [see Fig.~\ref{fig:Fig2}(b)]. In this particular order, the \textsf{MWIS}s correspond to the constraint satisfying assignments $(n_A,n_B,n_C)$ given by $(1,1,1)$, $(1,0,0)$, $(0,1,0)$ and $(0,0,1)$, respectively.\par
The \textsf{MWIS} encoding of Eq.~\eqref{eq:3MWIS} can also be directly used to construct the \textsf{MWIS} encoding of the four-body parity constraint. To do so one can merge two 3\textsf{MWIS} graphs by joining two nodes at the corners of the triangular 3\textsf{MWIS} graph depicted in Fig.~\ref{fig:Fig2}(b) and add the weights of the joint nodes [see Fig.~\ref{fig:Fig1}(c)]. Thus, the corresponding \textsf{MWIS}s have weight $6\alpha$ and correspond to the desired constraint-satisfying assignments $(n_A,n_B,n_C,n_D)$. As a result, combining the \textsf{MWIS} building blocks described above with the problem defining local fields on the parity-nodes realizes the modular \textsf{MWIS} formulation of the parity architecture [cf.~Eq.~\eqref{eq:localfield_hamiltonian} and Fig.~\ref{fig:Fig1}(b-c)].\par
\textit{Implementation including long-ranged tails.---} So far we have described the required parity constraints as \textsf{MWIS} problems on disk graphs where a proper choice of weights suffices to ensure a faithful representation of the problem. However, for a physical implementation on neutral atom hardware one has to take into account, that vdW interactions do not provide ``sharply-bounded'' discs but rather decay with inter-atomic distance $x$ as $1/x^6$ and therefore produce long-ranged interaction tails. As a consequence, this induces state-dependent biases, i.e., energy-differences between constraint satisfying configurations. In the following we describe a scalable construction recipe that is capable of systematically mitigating these errors and is based on ``modules'' and ``links'' connecting these modules [see Fig.~\ref{fig:Fig3}]. Each module consists of either a triangular or rhombic structure [see Fig.~\ref{fig:Fig2}], where the distance $d$ between neighbouring atoms is chosen as $d<r_B<\sqrt{3}d$. The rhombic structure appears whenever four-body constraints lie on top of each other such that sides of adjacent triangles can be merged as illustrated in Fig.~\ref{fig:Fig3}(b). In order to avoid inter-module crosstalk antiferromagnetically ordered links consisting of an odd number of atoms are used as a ``copying mechanism'' between connected module corners [see Fig.~\ref{fig:Fig3}(a)]. Furthermore, additional detunings on auxiliary qubits [see Fig.~\ref{fig:Fig3}(c)] are used for systematic compensation of biases on the module level (for details see Appendix~\ref{sec:Compensation}).\par
\begin{figure}[t]
\begin{centering}
\includegraphics[width =\columnwidth]{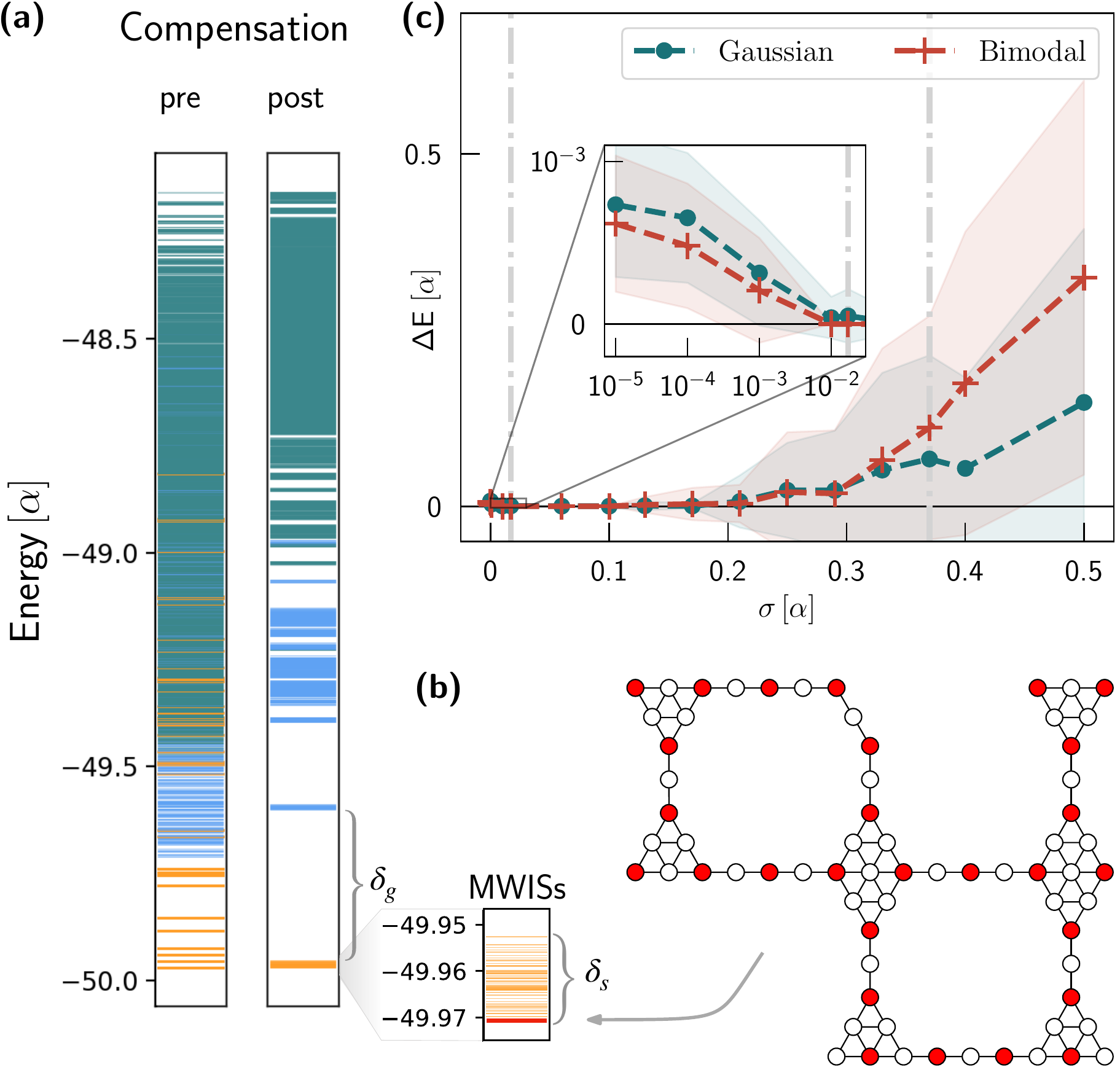}
\par\end{centering}
\protect\caption{\textit{Simulations.} 
(a) Energy spectrum of Hamiltonian Eq.~\eqref{eq:RydbergHamiltonian} before and after compensation of long-ranged interaction tails for the layout shown in Fig.~\ref{fig:Fig3}(a), without problem-defining local fields. The \textsf{MWIS} states with maximal weight $W_\textsf{max}$ are shown in orange. States belonging to independent sets of weight $(W_\textsf{max}-1)$ and weight $(W_\textsf{max}-2)$ are depicted in blue and green, respectively. 
(b) The lowest-energy \textsf{MWIS} configuration after compensation.
(c) Expected energy difference between the physical ground state and the state encoding the solution of the problem defined by local fields $J_m$, on the layout depicted in Fig.~\ref{fig:Fig3}(a).
The local fields are either Gaussian distributed with variance $\sigma^2$ or chosen from a bimodal distribution $\{\pm \sigma\}$. The vertical lines show the spread of the \textsf{MWIS} manifold ($\delta_\text{s}=0.017\alpha$) and the gap to the bulk ($\delta_\text{g} = 0.37\alpha$), respectively, for the scenario without local fields. Simulations are made for an interaction strength such that the nearest-neighbor interaction has a value of ${V(d) = 8 \alpha}$, where $d$ is the distance between two neighbouring atoms. 
}
\label{fig:Fig4}
\end{figure}
Without compensation, the states with maximal weight $W_\textsf{max}$ are neither energetically separated from the bulk, i.e., independent-set states with sub-optimal weight, nor energetically confined [see left panel of Fig.~\ref{fig:Fig4}(a)]. Note that as long as the nearest-neighbour interaction strength $V(d)$ is chosen sufficiently large with respect to $\alpha$, the blockade-violating states are well separated from the relevant $\textsf{MWIS}$ states. 
Our module-level compensation routine allows to efficiently minimize the energy spread $\delta_s$ within the \textsf{MWIS} manifold (for details see Appendix~\ref{sec:Compensation}) and thus leading to a clear separation to the bulk [see right panel of Fig.~\ref{fig:Fig4}(a)]. We want to emphasize that efficient compensation is also possible on extended parts of the layout comprised of several modules and links, which may even further improve the quality of compensation.\par
Together with proper normalization of problem-defining local fields (see Appendix~\ref{sec:Normalization}) this allows one to faithfully encode optimization problems in \textsf{MWIS} ground states of neutral atoms [see Fig.~\ref{fig:Fig4}(c)].
Specifically, if the fields are too weak, they cannot be resolved within the spread of the \textsf{MWIS} manifold. On the other hand, if the local fields are too strong, \textsf{MWIS} states begin to interfere with the bulk.\par
While the described scheme focuses on atoms arranged in a two-dimensional lattice geometry, we want to emphasize that a three-dimensional geometry would allow for further reduction in qubit numbers (see Appendix~\ref{sec:Simplifications}).

\textit{Conclusion and outlook.---}
In this work, we introduced a practically scalable scheme that allows for quantum optimization of higher-order constrained binary optimization problems readily realizable in state-of-the-art Rydberg devices. 
%
This is achieved by combining the advantages of the natural \textsf{MWIS} implementation in neutral atoms with the encoding advantages provided by the parity architecture.
Specifically, we introduced an implementation of the building blocks of the parity architecture, i.e., the three- and four-body constraints, as \textsf{MWIS} modules, including a local and modular compensation scheme dealing with long-ranged interaction tails originating from physical van der Waals interactions among qubits.\par
While this work focuses on faithfully encoding optimization problems, future work will address algorithmic aspects. In particular, our approach could be potentially used for programmable state preparation~\cite{Sieberer2018, Dlaska2019} and non-standard implementations of quantum approximate optimization algorithms~\cite{Farhi2014, Ender2022}.


\section*{Acknowledgements}
The authors thank Hannes Pichler for useful discussions. The \textsf{MWIS} problem instances were computed using the Julia package \href{https://github.com/QuEraComputing/GenericTensorNetworks.jl}{GenericTensorNetworks.jl}, which implements Ref.~\cite{Jin-Guo2022}. Work was supported by the Austrian Science Fund (FWF) through a START grant under Project No. Y1067-N27 and the SFB BeyondC Project No. F7108-N38. This project was funded within the QuantERA II Programme that has received funding from the European Union’s Horizon 2020 research and innovation programme under Grant Agreement No 101017733. \\

\textit{Note added.---} In the process of writing this manuscript we became aware of related work~\cite{Nguyen2022}. 


\appendix
\section{Compensation}
\label{sec:Compensation}

Our Rydberg-based approach to solving optimization problems presented in a ``parity format'' can be roughly divided into two steps.
In the first step, the parity constraints are reformulated as a \textsf{MWIS} problem. 
Here a proper choice of weights ensures a faithful representation of the problem.
In the second step, the \textsf{MWIS} problem is implemented on hardware by finding a unit disc configuration of the corresponding interaction graph. 
Due to the long-range Rydberg interactions, this implementation induces state-dependent biases such that different constraint-satisfying configurations are implemented by states having slightly different energies. 

In this section, we address this problem from a single three-body, and a four-body perspective and 
demonstrate a strategy to mitigate it. \\

In the case of a parity clause in three variables, we imagine the atoms being arranged in an equilateral triangle with base $2 d$, as depicted in Fig.~\ref{fig:Fig2}.
The first contribution $-\sum_i \Delta_i n_i$ in
Eq.~\eqref{eq:RydbergHamiltonian} does not distinguish between the four different \textsf{MWIS}s. However, the vdW contribution $\sum_{i<j}V(\boldsymbol{r_i})\hat{n}_i\hat{n}_v$ varies from state to state. Henceforth, we want to label a state by its defining \textsf{MWIS}. For example, 
the maximal set $\{A,B,C\}$ is physically represented by a state on 6 atoms which we denote by $|ABC\rangle$.
The energy difference between states $|ABC\rangle$ and $|Aa\rangle$
is given by 
\begin{equation}
 c_a = C_6\left(3 (2 d)^{-6} - (\sqrt{3} d)^{-6}\right). 
 \label{eq:3bodyCompensation}
\end{equation}
As the states $|ABC\rangle$ and $|Aa\rangle$ can be distinguished by $\hat{n}_a$, the replacement $\Delta_a \mapsto \Delta_a +  c_a$ compensates this geometry induced shift. Introducing the same compensation shifts $c_b = c_c = c_a$, all physical states corresponding to the four \textsf{MWIS}s are perfectly degenerate in energy.

\begin{figure}
    \centering
    \includegraphics[width=\columnwidth]{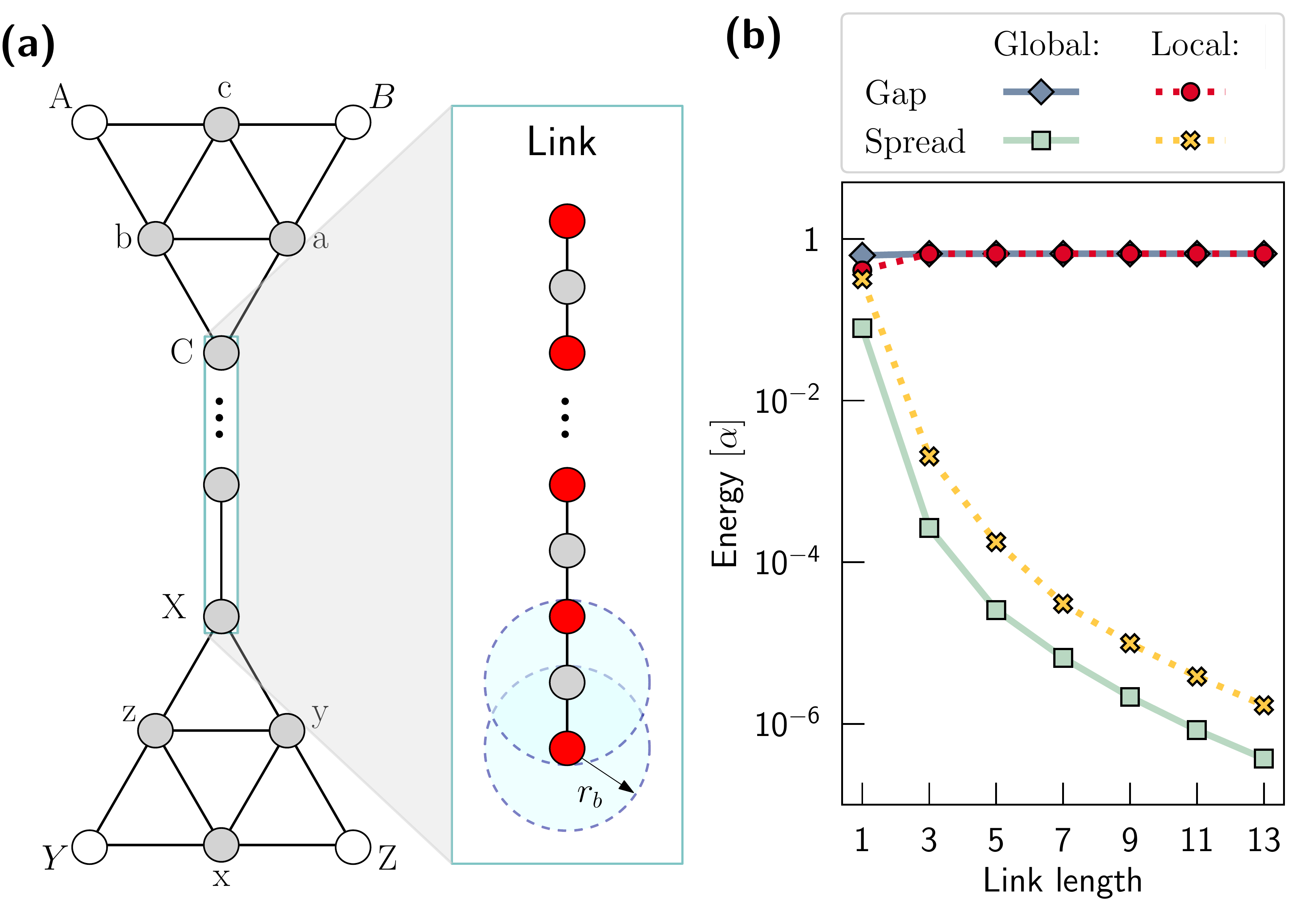}
    \caption{\textit{Compensation efficiency.} (a) A four-body parity constraint can be implemented by means of  basic three and two body building blocks [cf.Fig.~\ref{fig:Fig2}]. (b) Performance of two different compensation schemes in function of the length of the link. The ``Global'' column showcases the result for the optimal compensation scheme (see Sec.~\ref{sec:Compensation:4body}) when the system is viewed as a whole, while the ``Local'' column shows results for  cross-use of locally obtained compensation values (see Sec.~\ref{sec:Compensation:LocalCross}). The figure of merit depicted is the energy spread within states obeying the parity clause Eq.~\eqref{eq:BinaryConstraint} and the gap to non-fulfilling states. Simulations are made for an interaction strength such that the nearest neighbor interaction has a value of ${V(d) = 8 \alpha}$.
    }
    \label{fig:Fig5}
\end{figure}

Next, we discuss the four-body case. As described in the main text, the four body clause is implemented by means of two three-body clauses Eq.~\eqref{eq:2bodyBinaryConstraint}. We analyze the arrangement depicted in Fig.~\ref{fig:Fig4}(a). Built from basic building blocks [cf. Fig.~\ref{fig:Fig2}], two tree-body modules are connected via a link of an odd number of atoms. 
In the extreme case of an infinitely long link, module-module (MM) interactions vanish such that the analysis can be reduced to module-link interactions. It's illustrative to study the case of a single module with a link attached to it first. 
Let ABC denote the three-body module with a link attached to node C. Atoms within the links are placed at the same distance $d$ as nearest neighbours in the module.
Analogously to the free module case, compensation shifts can be obtained by balancing the energies of all four \textsf{MWIS}s. However, in contrast to the free case, the link breaks the rotational symmetry such that compensation energies no longer are symmetrical,
$\tilde{c}_a = \tilde{c}_b \neq \tilde{c}_c$.
Nevertheless, after compensation, the four \textsf{MWIS} states are perfectly degenerate in energy. To be more precise, these four states are given by the configuration of the three-body module with a unique continuation along the link. From the reference state of the single three-body module we construct the corresponding reference state $|ABC,\mathrm{link}\rangle$ with energy $\tilde{E}_r$. Likewise, we define $\tilde{E}_a$ as the vdW energy of the state $|Aa,\mathrm{link}\rangle$ and  $\tilde{E}_b$, $\tilde{E}_c$ analogously. Then, the compensation shifts are given by
\begin{equation}
        \tilde{c}_a = \tilde{E}_a - \tilde{E}_r, \quad
    \tilde{c}_b = \tilde{E}_b - \tilde{E}_r, \quad
    \tilde{c}_c = \tilde{E}_c - \tilde{E}_r .
\label{eq:MLcompensation}
\end{equation}

Things change when a second three-body module XYZ is introduced [see Fig.~\ref{fig:Fig5}(a)]. 
Imposing compensation on auxiliary nodes $a,b,c,$ and $x,y,z$  mitigates the vdW shifts only partiality as depicted in Fig.~\ref{fig:Fig4}(b). This is due to the non-vanishing MM interactions since \textsf{MWIS} states, which are indistinguishable on the first module may still differ on the second module. 
For example, the reference state $\ket{\mathrm{ref}}$ depicted in Fig.~\ref{fig:Fig5}(b) cannot be distinguished from configuration $\ket{x}$ by only looking at the upper module. On the other hand, link-module compensation is perfect as the information content of the link is locally accessible within the module. 

\begin{figure*}
    \centering
    \includegraphics[width=\textwidth]{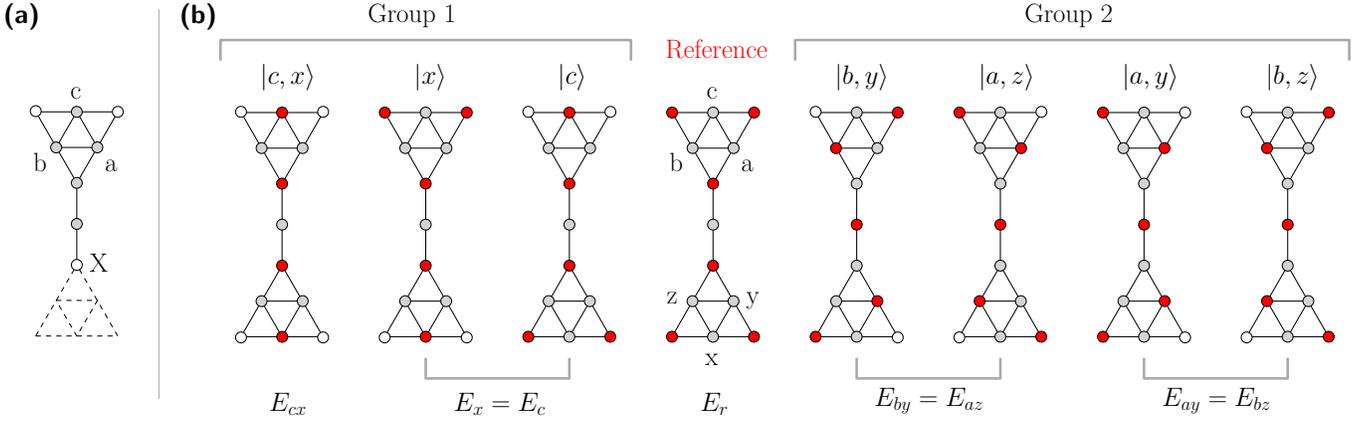}
    \caption{\textit{Relevant states for compensation of a four-body module.} (a) Local module-link compensation acts on one three-body module by taking into account the configuration of the link. (b) Global MLM compensation takes into account the full parity clause fulfilling subspace. The eight states are collected into three groups: A reference state and two groups which are decoupled from the action of local detunings.}
    \label{fig:Fig6}
\end{figure*}

In the following, we provide an optimal compensation scheme for the four-body case. ``Optimal'' in this context means that we look at the module-link-module (MLM) system as a whole and at all possible locally available degrees of freedom for compensation. Furthermore, we focus on reducing the energy spread within the \textsf{MWIS} manifold and neglect the impact of our changes on the gap between \textsf{MWIS} states and the bulk. This makes sense in the regime where the next nearest-neighbor interaction is not of the order $\alpha$.
In Sec.~\ref{sec:Compensation:LocalCross} we compare this result to an optimal compensation at module-link level. That means, we look at module one plus link (M1L) and module two plus link (M2L) individually, compensate them (which can be done perfectly), and use these compensation values in an adapted form for the whole MLM system. 

\subsection{Optimal four-body compensation}
\label{sec:Compensation:4body}
The MLM system exhibits eight different \textsf{MWIS} states.  These states are shown in Fig.~\ref{fig:Fig5}(b) for the special case of a link of length three. Here, we define the link as the atoms between the modules \emph{plus} the corners of the modules.  By this convention, overlapping modules have a link length of one. 

The eight \textsf{MWIS} states  can be collected into three groups: Group 0 consists of a single reference state, in which none of the auxiliary atoms $a,b,c,x,y,z$ is in the Rydberg state. Group 1 is a collection of three states, which can be influenced by local changes on auxiliary qubits $S_1 = \{c,x\}$. The remaining four states are influenced by local changes made on atoms $S_2 = \{a,b,y,z\}$.
To minimize the total energy spread, local detunings are introduced. These detunings are added on the auxiliary qubits with the aim of bringing all states in energy as close as possible to a
reference state with energy $E_r$ [see Fig.~\ref{fig:Fig5}(b)]. The reference state  $|\mathrm{ref}\rangle$ itself is unaffected by these detunings. Since the sets $S_1$ and $S_2$ are disjoint, both groups can be treated independently.

Let us discuss Group 1 first. Focus for a moment on the sites 'c' and 'x' [cf.~\ref{fig:Fig4}(a)]. Both of these sites can be either in the ground state $|g\rangle$ or in the Rydberg state $|r\rangle $. If we include the reference state, all four possible assignments are covered. From this local point of view we can identify the states in Group 1 as $|c\rangle := |rg\rangle_{cx}$, $|x\rangle := |gr\rangle_{cx}$ and 
$|cx\rangle := |rr\rangle_{cx}$.
Due to the symmetric arrangement the vdW energies of states $|c\rangle$ and $|x\rangle$ are the same. We will denote this energy as $E := E_c = E_x$.
Local detunings at sites $c$ and $x$ are imposed to bring all group 1 states energetically as close as possible to the reference state. This
can be formalized as
\begin{equation}
    \begin{split}
    K_1 := \min_{c_{c},c_x} \max( 
     &|E+c_c - E_r|, \\
     & |E+c_x - E_r|, \\ 
     & |E_{cx}+c_e+ c_x - E_r|),
\end{split}
\label{eq:Group1_Opt}
\end{equation}
where $E_r$ and $E_{c,x}$ denote the vdW energy of the reference state and the state $|cx\rangle$ respectively. 
By rescaling $x_1 :=  c_c/(E - E_r)$, $x_2 := c_{cx}/(E_{cx}-E_r)$ and introducing $R := (E_{cx}-E_r)/(E -E_r)$, Eq.~\eqref{eq:Group1_Opt} can be brought into a standard form
\begin{equation}
   K_1(R) := \min_{x_1,x_2} \max(|1+x_1|,|1+x_2|,|R+x_1+x_2|)
   \label{eq:Group1_StandardForm}.
\end{equation}
This problem can be rewritten as a linear program if one introduces a third variable $x_3$.  Terms like $|1+x_1|$ are replaced by two inequalities $1+x_1 \leq x_3$ and $-1-x_1 \leq x_3$. Consequently, Eq.~\eqref{eq:Group1_StandardForm} is equivalent to the problem of finding a vector $\boldsymbol{x} = (x_1,x_2,x_3)$ that maximizes $x_3$, subject to $A\boldsymbol{x} \leq \boldsymbol{b}$, where
\begin{equation}
A^t = 
\begin{bmatrix}
1 & -1 & 0 & 0 & 1 & -1\\
0 & 0 & 1 & -1 & 1 & -1\\
-1 & -1 & -1 & -1 & -1 & -1
\end{bmatrix}
\end{equation}
and $\boldsymbol{b}^t = [-1,1,-1,1,-R,R]$. The optimal solution to this linear program is given by
\begin{equation}
    x_1 = x_2 = -\frac{R+1}{3}
\end{equation}
with cost $x_3 = |1-R|/2$. 
Consequently,  the optimal compensation shifts are symmetrically given by
\begin{equation}
   c_c = c_{x} = -\frac{1}{3}\left(E +E_{cx} - 2 E_r \right).
\end{equation}

Next, we analyze Group 2. By symmetry, ${E_{by} = E_{az} =: E_1}$
and ${E_{ay} = E_{bz} =: E_2}$ holds. The four states are pushed energy-wise toward the reference state by minimizing
\begin{equation}
    \begin{split}
    K_2 := \min_{c_a,c_b,c_y,c_z} \max( 
     &|E_1 + c_{a} + c_{z} - E_r|, \\
     &|E_1 + c_{b} + c_{y} - E_r|, \\
     &|E_2 + c_{a} + c_{y} - E_r|, \\
     &|E_2 + c_{b} + c_{z} - E_r| ).
\end{split}
\label{eq:Group2_Opt}
\end{equation}
We introduce the shifted variables $E_1' := E_1 - E_r$, $E_2' := E_2 - E_r$ and abbreviate $R:=E_2'/E_1'$ Moreover, we introduce the variables  $x_1 := c_a/E_1',x_2 := c_b/E_1',x_3 := c_y/E_1'$ and $x_4 := c_z/E_1'$. Consequentially, Eq.~\eqref{eq:Group2_Opt} transforms into
\begin{equation}
       \begin{split}
     K_2(R) := \min_{\boldsymbol{x}} \max(&|1+x_1+x_4|,|1+x_2+x_3|, \\
     &|R+x_1+x_3|,|R+x_2+x_4|) .
\end{split}
   \label{eq:Group2_StandardForm}
\end{equation}
By solving the corresponding linear program, one gets the optimal basic feasible solutions 
\begin{align}
    \boldsymbol{x}_1^* &= -\frac{1}{2}[R+1,0,R+1,0]^t,\\
    \boldsymbol{x}_2^* &= -\frac{1}{2}[0,R+1,0,R+1)]^t .
\end{align}
For the numerical study presented in Fig.~\ref{fig:Fig4}(b) we choose a particular symmetric solution 
$\boldsymbol{x}^* := (\boldsymbol{x}_1^* + \boldsymbol{x}_2^*)/2$, where all compensation strengths are the same
\begin{equation}
    c_a=c_b=c_y=c_z =  -\frac{1}{4}\left(E_{1}+E_{2} - 2 E_r \right).
 \label{eq:4bodyCompensation_abyz}
\end{equation}
We would like to point out that adding additional detunings along the link cannot further reduce the spread, i.e., bringing all states closer to the reference state in energy. This is due to the fact that the link pattern is the same for all states in Group 1 given by $|rgr...gr\rangle$. Therefore, all these states would be affected in the same way by adding local detunings. The same argument holds for states in Group 2, where the link pattern is always of the form $|grg...rg\rangle$.
However, this discussion neglects the effect of compensation onto the gap. As already mentioned, this effect is assumed to be negligible when the nearest-neighbor interaction is not of order $\alpha$. Nonetheless, if we enter a regime where the nearest neighbor interactions are of order $\alpha$, it might be advantageous to include atoms along the link in the compensation process. 

\subsection{Local cross-compensation}
\label{sec:Compensation:LocalCross}
In this part we demonstrate the effectiveness of a local, module-link-based compensation scheme and compare it to the optimal compensation presented in the last section. This is again exemplarily done on a system consisting of two three-body modules connected by a link. 
This strategy can be thought of as calculating compensation values on the system ``module plus link''  and neglecting MM interactions.
Figure~\ref{fig:Fig5}(a)
displays the ML arrangement in the simple case of a link with length three. Note that atom `X' is considered to be part of the link as in any \textsf{MWIS} state atom X has a well-defined configuration after fixing a configuration of module one (M1). As mentioned already, M1 plus L can be compensated perfectly, giving rise to compensation values $\tilde{c}_a = \tilde{c}_b$ and $\tilde{c}_c$. By symmetry, M2 plus link can be compensated with the same values, i.e., $\tilde{c}_x = \tilde{c}_c$,  $\tilde{c}_y = \tilde{c}_a$ and  $\tilde{c}_z = \tilde{c}_b$. 
As MM interactions decrease with link length, it is expected that 
the compensation values obtained by considering ML interactions only, converge towards the full module-link-module (MLM) compensation values.
However, it turns out that, some ML values have to be slightly modified before they can be used in the MLM setting. 
This issue arises from the fact that we consider links with of odd length.  In that case, the two different link configurations which are compatible with blockade are
\begin{align}
    |\mathrm{link}_1\rangle &= |rgr...gr\rangle, \\
    |\mathrm{link}_2\rangle &= |grg...rg\rangle.
\end{align}
These states differ by one in the number of atoms in the Rydberg state.
The key insight is that states from Group 1 do have the same link pattern as the reference state having pattern $|\mathrm{link}_1\rangle$, while states from Group 2 don't. 
Hence, when calculating the optimal MLM compensation via Eq.~ \eqref{eq:Group1_Opt} energies $E-E_r$ have only contributions from module link interactions as the inter-link Rydberg contribution cancels out. 
From a local perspective, looking only at the module one without the link, the state $|c\rangle$ is recognized by $|Cc,\mathrm{link}\rangle$ while from the perspective of modules two  the same state is recognized as  $|\mathrm{Ref},\mathrm{link}\rangle$. In contrast, the state $|cx\rangle$ looks from both modules the same, namely as state $|Cc,\mathrm{link}\rangle$. Therefore, in the extreme of long links $(l\to \infty)$ Eq.~\eqref{eq:MLcompensation} reduces to
\begin{equation}
    \begin{split}
    c_c &= -\frac{1}{3}(E_c+E_{cr}-2E_r) \\
    & \to -\frac{1}{3}\left(\tilde{E}_c+\tilde{E}_r+2\tilde{E}_c-4\tilde{E}_r\right) \\
    &= -\left(\tilde{E}_c - \tilde{E}_r\right) = \tilde{c}_c
\end{split}
\end{equation}
Thus, for long links, one can efficiently use the ML compensation values via $\tilde{c}_c \approx c_c$ and $\tilde{c}_x \approx c_x$. 

However, states of Group 2 have the opposite link pattern $|\mathrm{link}_2\rangle$ compared to the reference state. When computing $E_1-E_r$ and $E_2-E_r$ in Eq.~\eqref{eq:Group2_Opt} this mismatch has to be taken into account. Henceforth, let 
\begin{equation}
    \aleph_l: = \frac{C_6}{d^6}\sum_{k=1}^{(l-1)/2}\frac{1}{(2k)^6}
\end{equation}
be the energy difference between state $|\mathrm{link}_1\rangle$ and $|\mathrm{link}_2\rangle$ for a link of length $l$. In the limit of an infinite long link, this additional energy shift converges to $\aleph_{\infty}  =C_6/(2d)^6\pi^6/945$.
To relate the local ML compensation values to the optimal MLM once, it is key to investigate states from Group 2 from a local perspective. Exemplary, state $|a,y\rangle$ does look from one side like state $|Aa,\mathrm{link}\rangle$ and from the other side like state $|Bb,\mathrm{link}\rangle$, both having the same energy $\tilde{E}$. The same considerations can be made for the remaining three states of Group 2. 
Hence, Eq.~\eqref{eq:4bodyCompensation_abyz} reduces in the infinite $(l\to \infty)$ limit to
\begin{equation}
    \begin{split}
    c_a = & -\frac{1}{4}(E_{ay}+E_{az}-2E_r) \\
    & \to -\frac{1}{4}(4 \tilde{E}_{a}+4\tilde{E}_{r}-2 \aleph_{\infty})
    = \tilde{c}_a - \frac{\aleph_{\infty}}{2}
\end{split}. 
\end{equation} 
Therefore, if the link is long enough 
$c_a \approx \tilde{c}_a - \aleph_{\infty}/2$ is the best local estimate. Figure~\ref{fig:Fig4}(b) shows the performance of the local ML compensation compared to the optimal global MLM  strategy.

\subsection{Generic module-link compensation}
\label{sec:Compensation:Generic}
The presented module-based compensation strategy has to be adjusted when more modules are involved. The noticeable difference from the situation discussed so far is that modules generally have multiple outgoing links or ``arms''. We exemplarily focus on the module shown in Fig.~\ref{fig:Fig3}(c), which has two arms. ML compensation nodes are labeled from $a$ to $c$. The reference state is uniquely defined as the \textsf{MWIS} where none of them is in the Rydberg state. That pattern uniquely extends into the two arms according to the pattern $|rgrg...\rangle$. Likewise, the remaining three \textsf{MWIS} states are given by a single excitation on the compensation nodes, i.e., $|a\rangle$,$|b\rangle$ and $|c\rangle$. By symmetry, states $|b\rangle$ and $|c\rangle$ do have link patterns that are in sync with the reference state in one arm, while being off-sync in the other arm, i.e., follow the pattern $|grgrg...\rangle$. As discussed before, that mismatch has to be compensated by an additional shift $- \aleph_{\infty}/2$. However, state  $|a\rangle$ is off-sync in both arms. Proper accounting is done by adding twice the amount, i.e., $- \aleph_{\infty}$, to the compensation shift $c_a$.

\section{Normalization}
\label{sec:Normalization}

As the optimization problem is mapped onto the \textsf{MWIS} problem, additional overhead is introduced. For faithful mapping, it has to be ensured that the \textsf{MWIS} encodes the solution to the original optimization problem. 
This requirement imposes a restriction on the magnitude of the weights $J_m$ in relation to $\alpha$, the scale of the unweighted problem.


At this scale, the MIWSs are separated from single defect configurations.
Such a single defect can manifest itself in one of our basic building blocks, either as a broken link or a broken three-body module. Exemplarily, for a single three-body module, these sub-optimal sets of weight $2\alpha$ are $\{A,B\}$, $\{A,C\}$, $\{B,C\}$ and the single element sets $\{a\},\{b\}$ and $\{c\}$.
From a conceptual point of view, a single defect configuration corresponds to a violation of one of the parity clauses.

In the special case of all-to-all connected two-body Ising problems, this issue was studied in detail in Ref.~\cite{Lanthaler2021}.
In that work, the authors investigate a
direct physical implementation of Eq.~\eqref{eq:localfield_hamiltonian} as a spin-model. As originally introduced in Ref.~\cite{Lechner2015}, the parity clauses Eq.~\eqref{eq:BinaryConstraint} are physically implemented via three and four-body terms
$\hat{\sigma}_z^1\hat{\sigma}_z^2\hat{\sigma}_z^3[\hat{\sigma}_z^4]$ imposing a energy penalty $C$ for constraint violating states.
That constraint strength is directly related to the \textsf{MWIS} scale $\alpha$. The analysis done in Ref.~\cite{Lanthaler2021} is thus directly applicable to the rescaling issue in the \textsf{MWIS} context.
The trivial and conservative lower bound is given by 
\begin{equation}
    \alpha \geq \sum_{ij}|J_{ij}|,
    \label{eq:TrivialAlphaLowerBound}
\end{equation}
which ensures a separation of the zero defect spectrum from the bulk. However, in most instances, this is a too strong requirement and even counterproductive for performing quantum annealing. There, the penalties should be neither too small nor too large. Analyzing several classes of fully connected optimization problems  Ref.~\cite{Lanthaler2021} found a sweet spot at which the fidelity (success probability) is maximized. 
These optimal strengths are defined by a gap condition, i.e., the parity model should faithfully represent the  ground state and first excited logical state. In contrast to full isolation of the logical subspace -- achieved by high penalties -- lowered penalties do not dominate over local fields encoding the problem. 
In the extreme case, for max-cut~\cite{Karp1972} on dense graphs, the scaling of constraints has to be observed to scale quadratic $\mathcal{O}(n^2)$ in system size. However, for more interesting classes, like the Sherrington-Kirkpatrick spin glass model~\cite{Parisi_1980} a linear increase $\mathcal{O}(n)$ is sufficient on average. 
In the simulations presented in Fig.~\ref{fig:Fig3}(e) it can be seen that the energy difference between physical to logical groundstate increases (with respect to $\alpha$) whenever the fluctuations in the local fields $J_m$ are comparable to the magnitude of the gap. 


\section{Three-dimensional layout}
\label{sec:Simplifications}

\begin{figure*}
    \centering
    \includegraphics[width = \textwidth]{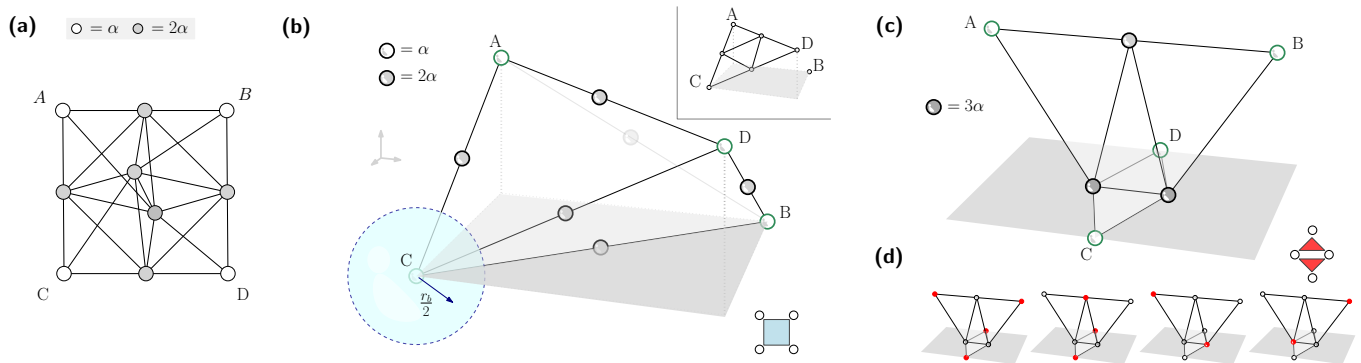}
    \caption{\textit{Three dimensional version of a four-body constraint and a twin three-body constraint.} Panel (a) shows a weighted graph on 10 nodes whose maximal sets represent the parity clause $n_A\oplus n_B \oplus n_C \oplus n_D = 0$. The graph from (a) can be represented by a unit-ball graph, as shown by the arrangement in panel (b). The inset showcases the possibility to implement a tree-body constraint by choosing one of the four sites of the tetraedron.
    In the case of two adjacent three-body modules (appearing whenever two four-body parity constraints are adjacent), a unit-ball \textsf{MWIS} representation on seven nodes is possible as shown in (c). (d) Displays the four relevant states.}
    \label{fig:Fig7}
\end{figure*}

In the main text, we have pursued the approach of building a four-body parity clause via Eq.~\eqref{eq:4bodyBinaryConstraint} from two three-body clauses. Section~\ref{sec:Compensation} demonstrates a compensation scheme to implement this approach in the most efficient way by spatially separating the three-body modules and connecting them via a link. However, the sub-optimal compensation is solely due to the fact that we stick to a 2D layout. Going to a 3D arrangement, a single four-body parity clause can be implemented with fewer atoms, and with perfect degeneracy. 
In the following, we will describe an alternative \textsf{MWIS} representation of a four-party clause on $10$ nodes [see Fig.~\ref{fig:Fig6}(a)]. This representation is not a unit-disc graph anymore but rather a 3D generalization of it. We will call this generalization a ``unit-ball graph''. Figure \ref{fig:Fig6}(b) showcases a possible 3D realization where the primary atoms are placed on the corners and the auxiliary ones along the edges of a tetrahedron. By adding detunings on the auxiliary atoms, the module can be compensated perfectly such that states corresponding to the eight \textsf{MWIS}s have the same energy. 
Going to three dimensions allows for further atom-saving ``hacks'' if the problem is built mainly from square plaquettes. Two adjacent three-body modules [see Fig.~\ref{fig:Fig3}(b)], as they would appear in two neighbouring four-body constraints] viewed as a single entity, permit a unit-ball implementation on seven atoms, as showcased in Fig.~\ref{fig:Fig6}(c-d). Despite experimental advances in trapping and arranging dozens of atoms arbitrarily arranged in space, it will be difficult to fully exploit these capabilities within a computational setting \cite{Barredo2018}.

\newpage

\begin{thebibliography}{28}%
\makeatletter
\providecommand \@ifxundefined [1]{%
 \@ifx{#1\undefined}
}%
\providecommand \@ifnum [1]{%
 \ifnum #1\expandafter \@firstoftwo
 \else \expandafter \@secondoftwo
 \fi
}%
\providecommand \@ifx [1]{%
 \ifx #1\expandafter \@firstoftwo
 \else \expandafter \@secondoftwo
 \fi
}%
\providecommand \natexlab [1]{#1}%
\providecommand \enquote  [1]{``#1''}%
\providecommand \bibnamefont  [1]{#1}%
\providecommand \bibfnamefont [1]{#1}%
\providecommand \citenamefont [1]{#1}%
\providecommand \href@noop [0]{\@secondoftwo}%
\providecommand \href [0]{\begingroup \@sanitize@url \@href}%
\providecommand \@href[1]{\@@startlink{#1}\@@href}%
\providecommand \@@href[1]{\endgroup#1\@@endlink}%
\providecommand \@sanitize@url [0]{\catcode `\\12\catcode `\$12\catcode
  `\&12\catcode `\#12\catcode `\^12\catcode `\_12\catcode `\%12\relax}%
\providecommand \@@startlink[1]{}%
\providecommand \@@endlink[0]{}%
\providecommand \url  [0]{\begingroup\@sanitize@url \@url }%
\providecommand \@url [1]{\endgroup\@href {#1}{\urlprefix }}%
\providecommand \urlprefix  [0]{URL }%
\providecommand \Eprint [0]{\href }%
\providecommand \doibase [0]{https://doi.org/}%
\providecommand \selectlanguage [0]{\@gobble}%
\providecommand \bibinfo  [0]{\@secondoftwo}%
\providecommand \bibfield  [0]{\@secondoftwo}%
\providecommand \translation [1]{[#1]}%
\providecommand \BibitemOpen [0]{}%
\providecommand \bibitemStop [0]{}%
\providecommand \bibitemNoStop [0]{.\EOS\space}%
\providecommand \EOS [0]{\spacefactor3000\relax}%
\providecommand \BibitemShut  [1]{\csname bibitem#1\endcsname}%
\let\auto@bib@innerbib\@empty
\bibitem [{\citenamefont {Saffman}\ \emph {et~al.}(2010)\citenamefont
  {Saffman}, \citenamefont {Walker},\ and\ \citenamefont
  {M\o{}lmer}}]{Saffman2010}%
  \BibitemOpen
  \bibfield  {author} {\bibinfo {author} {\bibfnamefont {M.}~\bibnamefont
  {Saffman}}, \bibinfo {author} {\bibfnamefont {T.~G.}\ \bibnamefont
  {Walker}},\ and\ \bibinfo {author} {\bibfnamefont {K.}~\bibnamefont
  {M\o{}lmer}},\ }\bibfield  {title} {\bibinfo {title} {Quantum information
  with {R}ydberg atoms},\ }\href {https://doi.org/10.1103/RevModPhys.82.2313}
  {\bibfield  {journal} {\bibinfo  {journal} {Rev. Mod. Phys.}\ }\textbf
  {\bibinfo {volume} {82}},\ \bibinfo {pages} {2313} (\bibinfo {year}
  {2010})}\BibitemShut {NoStop}%
\bibitem [{\citenamefont {Henriet}\ \emph {et~al.}(2020)\citenamefont
  {Henriet}, \citenamefont {Beguin}, \citenamefont {Signoles}, \citenamefont
  {Lahaye}, \citenamefont {Browaeys}, \citenamefont {Reymond},\ and\
  \citenamefont {Jurczak}}]{Henriet2020}%
  \BibitemOpen
  \bibfield  {author} {\bibinfo {author} {\bibfnamefont {L.}~\bibnamefont
  {Henriet}}, \bibinfo {author} {\bibfnamefont {L.}~\bibnamefont {Beguin}},
  \bibinfo {author} {\bibfnamefont {A.}~\bibnamefont {Signoles}}, \bibinfo
  {author} {\bibfnamefont {T.}~\bibnamefont {Lahaye}}, \bibinfo {author}
  {\bibfnamefont {A.}~\bibnamefont {Browaeys}}, \bibinfo {author}
  {\bibfnamefont {G.-O.}\ \bibnamefont {Reymond}},\ and\ \bibinfo {author}
  {\bibfnamefont {C.}~\bibnamefont {Jurczak}},\ }\bibfield  {title} {\bibinfo
  {title} {Quantum computing with neutral atoms},\ }\href
  {https://doi.org/10.22331/q-2020-09-21-327} {\bibfield  {journal} {\bibinfo
  {journal} {{Quantum}}\ }\textbf {\bibinfo {volume} {4}},\ \bibinfo {pages}
  {327} (\bibinfo {year} {2020})}\BibitemShut {NoStop}%
\bibitem [{\citenamefont {Scholl}\ \emph {et~al.}(2021)\citenamefont {Scholl},
  \citenamefont {Schuler}, \citenamefont {Williams}, \citenamefont
  {Eberharter}, \citenamefont {Barredo}, \citenamefont {Schymik}, \citenamefont
  {Lienhard}, \citenamefont {Henry}, \citenamefont {Lang}, \citenamefont
  {Lahaye}, \citenamefont {L{\"a}uchli},\ and\ \citenamefont
  {Browaeys}}]{Scholl2021}%
  \BibitemOpen
  \bibfield  {author} {\bibinfo {author} {\bibfnamefont {P.}~\bibnamefont
  {Scholl}}, \bibinfo {author} {\bibfnamefont {M.}~\bibnamefont {Schuler}},
  \bibinfo {author} {\bibfnamefont {H.~J.}\ \bibnamefont {Williams}}, \bibinfo
  {author} {\bibfnamefont {A.~A.}\ \bibnamefont {Eberharter}}, \bibinfo
  {author} {\bibfnamefont {D.}~\bibnamefont {Barredo}}, \bibinfo {author}
  {\bibfnamefont {K.-N.}\ \bibnamefont {Schymik}}, \bibinfo {author}
  {\bibfnamefont {V.}~\bibnamefont {Lienhard}}, \bibinfo {author}
  {\bibfnamefont {L.-P.}\ \bibnamefont {Henry}}, \bibinfo {author}
  {\bibfnamefont {T.~C.}\ \bibnamefont {Lang}}, \bibinfo {author}
  {\bibfnamefont {T.}~\bibnamefont {Lahaye}}, \bibinfo {author} {\bibfnamefont
  {A.~M.}\ \bibnamefont {L{\"a}uchli}},\ and\ \bibinfo {author} {\bibfnamefont
  {A.}~\bibnamefont {Browaeys}},\ }\bibfield  {title} {\bibinfo {title}
  {{Quantum simulation of 2D antiferromagnets with hundreds of Rydberg
  atoms}},\ }\href {https://doi.org/10.1038/s41586-021-03585-1} {\bibfield
  {journal} {\bibinfo  {journal} {Nature}\ }\textbf {\bibinfo {volume} {595}},\
  \bibinfo {pages} {233} (\bibinfo {year} {2021})}\BibitemShut {NoStop}%
\bibitem [{\citenamefont {Bluvstein}\ \emph {et~al.}(2021)\citenamefont
  {Bluvstein}, \citenamefont {Omran}, \citenamefont {Levine}, \citenamefont
  {Keesling}, \citenamefont {Semeghini}, \citenamefont {Ebadi}, \citenamefont
  {Wang}, \citenamefont {Michailidis}, \citenamefont {Maskara}, \citenamefont
  {Ho}, \citenamefont {Choi}, \citenamefont {Serbyn}, \citenamefont {Greiner},
  \citenamefont {Vuleti{\'c}},\ and\ \citenamefont {Lukin}}]{Bluvstein2021}%
  \BibitemOpen
  \bibfield  {author} {\bibinfo {author} {\bibfnamefont {D.}~\bibnamefont
  {Bluvstein}}, \bibinfo {author} {\bibfnamefont {A.}~\bibnamefont {Omran}},
  \bibinfo {author} {\bibfnamefont {H.}~\bibnamefont {Levine}}, \bibinfo
  {author} {\bibfnamefont {A.}~\bibnamefont {Keesling}}, \bibinfo {author}
  {\bibfnamefont {G.}~\bibnamefont {Semeghini}}, \bibinfo {author}
  {\bibfnamefont {S.}~\bibnamefont {Ebadi}}, \bibinfo {author} {\bibfnamefont
  {T.~T.}\ \bibnamefont {Wang}}, \bibinfo {author} {\bibfnamefont {A.~A.}\
  \bibnamefont {Michailidis}}, \bibinfo {author} {\bibfnamefont
  {N.}~\bibnamefont {Maskara}}, \bibinfo {author} {\bibfnamefont {W.~W.}\
  \bibnamefont {Ho}}, \bibinfo {author} {\bibfnamefont {S.}~\bibnamefont
  {Choi}}, \bibinfo {author} {\bibfnamefont {M.}~\bibnamefont {Serbyn}},
  \bibinfo {author} {\bibfnamefont {M.}~\bibnamefont {Greiner}}, \bibinfo
  {author} {\bibfnamefont {V.}~\bibnamefont {Vuleti{\'c}}},\ and\ \bibinfo
  {author} {\bibfnamefont {M.~D.}\ \bibnamefont {Lukin}},\ }\bibfield  {title}
  {\bibinfo {title} {{Controlling quantum many-body dynamics in driven
  {R}ydberg atom arrays}},\ }\href {https://doi.org/10.1126/science.abg2530}
  {\bibfield  {journal} {\bibinfo  {journal} {Science}\ }\textbf {\bibinfo
  {volume} {371}},\ \bibinfo {pages} {1355} (\bibinfo {year}
  {2021})}\BibitemShut {NoStop}%
\bibitem [{\citenamefont {Ebadi}\ \emph {et~al.}(2021)\citenamefont {Ebadi},
  \citenamefont {Wang}, \citenamefont {Levine}, \citenamefont {Keesling},
  \citenamefont {Semeghini}, \citenamefont {Omran}, \citenamefont {Bluvstein},
  \citenamefont {Samajdar}, \citenamefont {Pichler}, \citenamefont {Ho},
  \citenamefont {Choi}, \citenamefont {Sachdev}, \citenamefont {Greiner},
  \citenamefont {Vuleti{\'c}},\ and\ \citenamefont {Lukin}}]{Ebadi2021}%
  \BibitemOpen
  \bibfield  {author} {\bibinfo {author} {\bibfnamefont {S.}~\bibnamefont
  {Ebadi}}, \bibinfo {author} {\bibfnamefont {T.~T.}\ \bibnamefont {Wang}},
  \bibinfo {author} {\bibfnamefont {H.}~\bibnamefont {Levine}}, \bibinfo
  {author} {\bibfnamefont {A.}~\bibnamefont {Keesling}}, \bibinfo {author}
  {\bibfnamefont {G.}~\bibnamefont {Semeghini}}, \bibinfo {author}
  {\bibfnamefont {A.}~\bibnamefont {Omran}}, \bibinfo {author} {\bibfnamefont
  {D.}~\bibnamefont {Bluvstein}}, \bibinfo {author} {\bibfnamefont
  {R.}~\bibnamefont {Samajdar}}, \bibinfo {author} {\bibfnamefont
  {H.}~\bibnamefont {Pichler}}, \bibinfo {author} {\bibfnamefont {W.~W.}\
  \bibnamefont {Ho}}, \bibinfo {author} {\bibfnamefont {S.}~\bibnamefont
  {Choi}}, \bibinfo {author} {\bibfnamefont {S.}~\bibnamefont {Sachdev}},
  \bibinfo {author} {\bibfnamefont {M.}~\bibnamefont {Greiner}}, \bibinfo
  {author} {\bibfnamefont {V.}~\bibnamefont {Vuleti{\'c}}},\ and\ \bibinfo
  {author} {\bibfnamefont {M.~D.}\ \bibnamefont {Lukin}},\ }\bibfield  {title}
  {\bibinfo {title} {Quantum phases of matter on a 256-atom programmable
  quantum simulator},\ }\href {https://doi.org/10.1038/s41586-021-03582-4}
  {\bibfield  {journal} {\bibinfo  {journal} {Nature}\ }\textbf {\bibinfo
  {volume} {595}},\ \bibinfo {pages} {227} (\bibinfo {year}
  {2021})}\BibitemShut {NoStop}%
\bibitem [{\citenamefont {Ebadi}\ \emph {et~al.}(2022)\citenamefont {Ebadi},
  \citenamefont {Keesling}, \citenamefont {Cain}, \citenamefont {Wang},
  \citenamefont {Levine}, \citenamefont {Bluvstein}, \citenamefont {Semeghini},
  \citenamefont {Omran}, \citenamefont {Liu}, \citenamefont {Samajdar} \emph
  {et~al.}}]{Ebadi2022}%
  \BibitemOpen
  \bibfield  {author} {\bibinfo {author} {\bibfnamefont {S.}~\bibnamefont
  {Ebadi}}, \bibinfo {author} {\bibfnamefont {A.}~\bibnamefont {Keesling}},
  \bibinfo {author} {\bibfnamefont {M.}~\bibnamefont {Cain}}, \bibinfo {author}
  {\bibfnamefont {T.~T.}\ \bibnamefont {Wang}}, \bibinfo {author}
  {\bibfnamefont {H.}~\bibnamefont {Levine}}, \bibinfo {author} {\bibfnamefont
  {D.}~\bibnamefont {Bluvstein}}, \bibinfo {author} {\bibfnamefont
  {G.}~\bibnamefont {Semeghini}}, \bibinfo {author} {\bibfnamefont
  {A.}~\bibnamefont {Omran}}, \bibinfo {author} {\bibfnamefont {J.-G.}\
  \bibnamefont {Liu}}, \bibinfo {author} {\bibfnamefont {R.}~\bibnamefont
  {Samajdar}}, \emph {et~al.},\ }\bibfield  {title} {\bibinfo {title} {Quantum
  optimization of maximum independent set using {R}ydberg atom arrays},\ }\href
  {https://doi.org/10.1126/science.abo6587} {\bibfield  {journal} {\bibinfo
  {journal} {Science}\ }\textbf {\bibinfo {volume} {376}},\ \bibinfo {pages}
  {1209} (\bibinfo {year} {2022})}\BibitemShut {NoStop}%
\bibitem [{\citenamefont {Graham}\ \emph {et~al.}(2022)\citenamefont {Graham},
  \citenamefont {Song}, \citenamefont {Scott}, \citenamefont {Poole},
  \citenamefont {Phuttitarn}, \citenamefont {Jooya}, \citenamefont {Eichler},
  \citenamefont {Jiang}, \citenamefont {Marra}, \citenamefont {Grinkemeyer},
  \citenamefont {Kwon}, \citenamefont {Ebert}, \citenamefont {Cherek},
  \citenamefont {Lichtman}, \citenamefont {Gillette}, \citenamefont {Gilbert},
  \citenamefont {Bowman}, \citenamefont {Ballance}, \citenamefont {Campbell},
  \citenamefont {Dahl}, \citenamefont {Crawford}, \citenamefont {Blunt},
  \citenamefont {Rogers}, \citenamefont {Noel},\ and\ \citenamefont
  {Saffman}}]{Graham2022}%
  \BibitemOpen
  \bibfield  {author} {\bibinfo {author} {\bibfnamefont {T.~M.}\ \bibnamefont
  {Graham}}, \bibinfo {author} {\bibfnamefont {Y.}~\bibnamefont {Song}},
  \bibinfo {author} {\bibfnamefont {J.}~\bibnamefont {Scott}}, \bibinfo
  {author} {\bibfnamefont {C.}~\bibnamefont {Poole}}, \bibinfo {author}
  {\bibfnamefont {L.}~\bibnamefont {Phuttitarn}}, \bibinfo {author}
  {\bibfnamefont {K.}~\bibnamefont {Jooya}}, \bibinfo {author} {\bibfnamefont
  {P.}~\bibnamefont {Eichler}}, \bibinfo {author} {\bibfnamefont
  {X.}~\bibnamefont {Jiang}}, \bibinfo {author} {\bibfnamefont
  {A.}~\bibnamefont {Marra}}, \bibinfo {author} {\bibfnamefont
  {B.}~\bibnamefont {Grinkemeyer}}, \bibinfo {author} {\bibfnamefont
  {M.}~\bibnamefont {Kwon}}, \bibinfo {author} {\bibfnamefont {M.}~\bibnamefont
  {Ebert}}, \bibinfo {author} {\bibfnamefont {J.}~\bibnamefont {Cherek}},
  \bibinfo {author} {\bibfnamefont {M.~T.}\ \bibnamefont {Lichtman}}, \bibinfo
  {author} {\bibfnamefont {M.}~\bibnamefont {Gillette}}, \bibinfo {author}
  {\bibfnamefont {J.}~\bibnamefont {Gilbert}}, \bibinfo {author} {\bibfnamefont
  {D.}~\bibnamefont {Bowman}}, \bibinfo {author} {\bibfnamefont
  {T.}~\bibnamefont {Ballance}}, \bibinfo {author} {\bibfnamefont
  {C.}~\bibnamefont {Campbell}}, \bibinfo {author} {\bibfnamefont {E.~D.}\
  \bibnamefont {Dahl}}, \bibinfo {author} {\bibfnamefont {O.}~\bibnamefont
  {Crawford}}, \bibinfo {author} {\bibfnamefont {N.~S.}\ \bibnamefont {Blunt}},
  \bibinfo {author} {\bibfnamefont {B.}~\bibnamefont {Rogers}}, \bibinfo
  {author} {\bibfnamefont {T.}~\bibnamefont {Noel}},\ and\ \bibinfo {author}
  {\bibfnamefont {M.}~\bibnamefont {Saffman}},\ }\bibfield  {title} {\bibinfo
  {title} {Multi-qubit entanglement and algorithms on a neutral-atom quantum
  computer},\ }\href {https://doi.org/10.1038/s41586-022-04603-6} {\bibfield
  {journal} {\bibinfo  {journal} {Nature}\ }\textbf {\bibinfo {volume} {604}},\
  \bibinfo {pages} {457} (\bibinfo {year} {2022})}\BibitemShut {NoStop}%
\bibitem [{\citenamefont {Preskill}(2018)}]{Preskill2018_nisq}%
  \BibitemOpen
  \bibfield  {author} {\bibinfo {author} {\bibfnamefont {J.}~\bibnamefont
  {Preskill}},\ }\bibfield  {title} {\bibinfo {title} {Quantum computing in the
  nisq era and beyond},\ }\href {https://doi.org/10.22331/q-2018-08-06-79}
  {\bibfield  {journal} {\bibinfo  {journal} {Quantum}\ }\textbf {\bibinfo
  {volume} {2}},\ \bibinfo {pages} {79} (\bibinfo {year} {2018})}\BibitemShut
  {NoStop}%
\bibitem [{\citenamefont {Pichler}\ \emph
  {et~al.}(2018{\natexlab{a}})\citenamefont {Pichler}, \citenamefont {Wang},
  \citenamefont {Zhou}, \citenamefont {Choi},\ and\ \citenamefont
  {Lukin}}]{Pichler2018}%
  \BibitemOpen
  \bibfield  {author} {\bibinfo {author} {\bibfnamefont {H.}~\bibnamefont
  {Pichler}}, \bibinfo {author} {\bibfnamefont {S.-T.}\ \bibnamefont {Wang}},
  \bibinfo {author} {\bibfnamefont {L.}~\bibnamefont {Zhou}}, \bibinfo {author}
  {\bibfnamefont {S.}~\bibnamefont {Choi}},\ and\ \bibinfo {author}
  {\bibfnamefont {M.~D.}\ \bibnamefont {Lukin}},\ }\href@noop {} {\bibinfo
  {title} {Quantum optimization for maximum independent set using {R}ydberg
  atom arrays}} (\bibinfo {year} {2018}{\natexlab{a}}),\ \Eprint
  {https://arxiv.org/abs/1808.10816} {arXiv:1808.10816} \BibitemShut {NoStop}%
\bibitem [{\citenamefont {Jaksch}\ \emph {et~al.}(2000)\citenamefont {Jaksch},
  \citenamefont {Cirac}, \citenamefont {Zoller}, \citenamefont {Rolston},
  \citenamefont {C\^ot\'e},\ and\ \citenamefont {Lukin}}]{Jacksch2000}%
  \BibitemOpen
  \bibfield  {author} {\bibinfo {author} {\bibfnamefont {D.}~\bibnamefont
  {Jaksch}}, \bibinfo {author} {\bibfnamefont {J.~I.}\ \bibnamefont {Cirac}},
  \bibinfo {author} {\bibfnamefont {P.}~\bibnamefont {Zoller}}, \bibinfo
  {author} {\bibfnamefont {S.~L.}\ \bibnamefont {Rolston}}, \bibinfo {author}
  {\bibfnamefont {R.}~\bibnamefont {C\^ot\'e}},\ and\ \bibinfo {author}
  {\bibfnamefont {M.~D.}\ \bibnamefont {Lukin}},\ }\bibfield  {title} {\bibinfo
  {title} {Fast quantum gates for neutral atoms},\ }\href
  {https://doi.org/10.1103/PhysRevLett.85.2208} {\bibfield  {journal} {\bibinfo
   {journal} {Phys. Rev. Lett.}\ }\textbf {\bibinfo {volume} {85}},\ \bibinfo
  {pages} {2208} (\bibinfo {year} {2000})}\BibitemShut {NoStop}%
\bibitem [{\citenamefont {Lukin}\ \emph {et~al.}(2001)\citenamefont {Lukin},
  \citenamefont {Fleischhauer}, \citenamefont {Cote}, \citenamefont {Duan},
  \citenamefont {Jaksch}, \citenamefont {Cirac},\ and\ \citenamefont
  {Zoller}}]{Lukin2001}%
  \BibitemOpen
  \bibfield  {author} {\bibinfo {author} {\bibfnamefont {M.~D.}\ \bibnamefont
  {Lukin}}, \bibinfo {author} {\bibfnamefont {M.}~\bibnamefont {Fleischhauer}},
  \bibinfo {author} {\bibfnamefont {R.}~\bibnamefont {Cote}}, \bibinfo {author}
  {\bibfnamefont {L.~M.}\ \bibnamefont {Duan}}, \bibinfo {author}
  {\bibfnamefont {D.}~\bibnamefont {Jaksch}}, \bibinfo {author} {\bibfnamefont
  {J.~I.}\ \bibnamefont {Cirac}},\ and\ \bibinfo {author} {\bibfnamefont
  {P.}~\bibnamefont {Zoller}},\ }\bibfield  {title} {\bibinfo {title} {Dipole
  blockade and quantum information processing in mesoscopic atomic ensembles},\
  }\href {https://doi.org/10.1103/PhysRevLett.87.037901} {\bibfield  {journal}
  {\bibinfo  {journal} {Phys. Rev. Lett.}\ }\textbf {\bibinfo {volume} {87}},\
  \bibinfo {pages} {037901} (\bibinfo {year} {2001})}\BibitemShut {NoStop}%
\bibitem [{\citenamefont {Ga{\"e}tan}\ \emph {et~al.}(2009)\citenamefont
  {Ga{\"e}tan}, \citenamefont {Miroshnychenko}, \citenamefont {Wilk},
  \citenamefont {Chotia}, \citenamefont {Viteau}, \citenamefont {Comparat},
  \citenamefont {Pillet}, \citenamefont {Browaeys},\ and\ \citenamefont
  {Grangier}}]{Gaetan2009}%
  \BibitemOpen
  \bibfield  {author} {\bibinfo {author} {\bibfnamefont {A.}~\bibnamefont
  {Ga{\"e}tan}}, \bibinfo {author} {\bibfnamefont {Y.}~\bibnamefont
  {Miroshnychenko}}, \bibinfo {author} {\bibfnamefont {T.}~\bibnamefont
  {Wilk}}, \bibinfo {author} {\bibfnamefont {A.}~\bibnamefont {Chotia}},
  \bibinfo {author} {\bibfnamefont {M.}~\bibnamefont {Viteau}}, \bibinfo
  {author} {\bibfnamefont {D.}~\bibnamefont {Comparat}}, \bibinfo {author}
  {\bibfnamefont {P.}~\bibnamefont {Pillet}}, \bibinfo {author} {\bibfnamefont
  {A.}~\bibnamefont {Browaeys}},\ and\ \bibinfo {author} {\bibfnamefont
  {P.}~\bibnamefont {Grangier}},\ }\bibfield  {title} {\bibinfo {title}
  {Observation of collective excitation of two individual atoms in the
  {R}ydberg blockade regime},\ }\href {https://doi.org/10.1038/nphys1183}
  {\bibfield  {journal} {\bibinfo  {journal} {Nature Physics}\ }\textbf
  {\bibinfo {volume} {5}},\ \bibinfo {pages} {115} (\bibinfo {year}
  {2009})}\BibitemShut {NoStop}%
\bibitem [{\citenamefont {Pichler}\ \emph
  {et~al.}(2018{\natexlab{b}})\citenamefont {Pichler}, \citenamefont {Wang},
  \citenamefont {Zhou}, \citenamefont {Choi},\ and\ \citenamefont
  {Lukin}}]{Pichler2018_complexity}%
  \BibitemOpen
  \bibfield  {author} {\bibinfo {author} {\bibfnamefont {H.}~\bibnamefont
  {Pichler}}, \bibinfo {author} {\bibfnamefont {S.-T.}\ \bibnamefont {Wang}},
  \bibinfo {author} {\bibfnamefont {L.}~\bibnamefont {Zhou}}, \bibinfo {author}
  {\bibfnamefont {S.}~\bibnamefont {Choi}},\ and\ \bibinfo {author}
  {\bibfnamefont {M.~D.}\ \bibnamefont {Lukin}},\ }\href@noop {} {\bibinfo
  {title} {Computational complexity of the {R}ydberg blockade in two
  dimensions}} (\bibinfo {year} {2018}{\natexlab{b}}),\ \Eprint
  {https://arxiv.org/abs/1809.04954} {arXiv:1809.04954} \BibitemShut {NoStop}%
\bibitem [{\citenamefont {Lechner}\ \emph {et~al.}(2015)\citenamefont
  {Lechner}, \citenamefont {Hauke},\ and\ \citenamefont
  {Zoller}}]{Lechner2015}%
  \BibitemOpen
  \bibfield  {author} {\bibinfo {author} {\bibfnamefont {W.}~\bibnamefont
  {Lechner}}, \bibinfo {author} {\bibfnamefont {P.}~\bibnamefont {Hauke}},\
  and\ \bibinfo {author} {\bibfnamefont {P.}~\bibnamefont {Zoller}},\
  }\bibfield  {title} {\bibinfo {title} {A quantum annealing architecture with
  all-to-all connectivity from local interactions},\ }\href
  {https://advances.sciencemag.org/content/1/9/e1500838} {\bibfield  {journal}
  {\bibinfo  {journal} {Science Advances}\ }\textbf {\bibinfo {volume} {1}},\
  \bibinfo {pages} {e1500838} (\bibinfo {year} {2015})}\BibitemShut {NoStop}%
\bibitem [{\citenamefont {Ender}\ \emph {et~al.}(2021)\citenamefont {Ender},
  \citenamefont {ter Hoeven}, \citenamefont {Niehoff}, \citenamefont
  {Drieb-Schön},\ and\ \citenamefont {Lechner}}]{Ender2021}%
  \BibitemOpen
  \bibfield  {author} {\bibinfo {author} {\bibfnamefont {K.}~\bibnamefont
  {Ender}}, \bibinfo {author} {\bibfnamefont {R.}~\bibnamefont {ter Hoeven}},
  \bibinfo {author} {\bibfnamefont {B.~E.}\ \bibnamefont {Niehoff}}, \bibinfo
  {author} {\bibfnamefont {M.}~\bibnamefont {Drieb-Schön}},\ and\ \bibinfo
  {author} {\bibfnamefont {W.}~\bibnamefont {Lechner}},\ }\href@noop {}
  {\bibinfo {title} {Parity quantum optimization: Compiler}} (\bibinfo {year}
  {2021}),\ \Eprint {https://arxiv.org/abs/2105.06233} {arXiv:2105.06233}
  \BibitemShut {NoStop}%
\bibitem [{\citenamefont {Karp}(1972)}]{Karp1972}%
  \BibitemOpen
  \bibfield  {author} {\bibinfo {author} {\bibfnamefont {R.~M.}\ \bibnamefont
  {Karp}},\ }\bibinfo {title} {Reducibility among combinatorial problems},\ in\
  \href {https://doi.org/10.1007/978-1-4684-2001-2_9} {\emph {\bibinfo
  {booktitle} {Complexity of Computer Computations}}},\ \bibinfo {editor}
  {edited by\ \bibinfo {editor} {\bibfnamefont {R.~E.}\ \bibnamefont {Miller}},
  \bibinfo {editor} {\bibfnamefont {J.~W.}\ \bibnamefont {Thatcher}},\ and\
  \bibinfo {editor} {\bibfnamefont {J.~D.}\ \bibnamefont {Bohlinger}}}\
  (\bibinfo  {publisher} {Springer US},\ \bibinfo {address} {Boston, MA},\
  \bibinfo {year} {1972})\ pp.\ \bibinfo {pages} {85--103}\BibitemShut
  {NoStop}%
\bibitem [{\citenamefont {Fishkin}(2004)}]{Fishkin2004}%
  \BibitemOpen
  \bibfield  {author} {\bibinfo {author} {\bibfnamefont {A.~V.}\ \bibnamefont
  {Fishkin}},\ }\bibinfo {title} {Disk graphs: A short survey},\ in\ \href
  {https://doi.org/10.1007/978-3-540-24592-6_23} {\emph {\bibinfo {booktitle}
  {Approximation and Online Algorithms}}},\ \bibinfo {editor} {edited by\
  \bibinfo {editor} {\bibfnamefont {R.}~\bibnamefont {Solis-Oba}}\ and\
  \bibinfo {editor} {\bibfnamefont {K.}~\bibnamefont {Jansen}}}\ (\bibinfo
  {publisher} {Springer Berlin Heidelberg},\ \bibinfo {address} {Berlin,
  Heidelberg},\ \bibinfo {year} {2004})\ pp.\ \bibinfo {pages}
  {260--264}\BibitemShut {NoStop}%
\bibitem [{\citenamefont {Albash}\ and\ \citenamefont
  {Lidar}(2018)}]{Albash2018}%
  \BibitemOpen
  \bibfield  {author} {\bibinfo {author} {\bibfnamefont {T.}~\bibnamefont
  {Albash}}\ and\ \bibinfo {author} {\bibfnamefont {D.~A.}\ \bibnamefont
  {Lidar}},\ }\bibfield  {title} {\bibinfo {title} {Adiabatic quantum
  computation},\ }\href {https://link.aps.org/doi/10.1103/RevModPhys.90.015002}
  {\bibfield  {journal} {\bibinfo  {journal} {Reviews of Modern Physics}\
  }\textbf {\bibinfo {volume} {90}},\ \bibinfo {pages} {015002} (\bibinfo
  {year} {2018})}\BibitemShut {NoStop}%
\bibitem [{Note1()}]{Note1}%
  \BibitemOpen
  \bibinfo {note} {Note, that this is the minimum number of vertices which
  allow for such an encoding.}\BibitemShut {Stop}%
\bibitem [{\citenamefont {Sieberer}\ and\ \citenamefont
  {Lechner}(2018)}]{Sieberer2018}%
  \BibitemOpen
  \bibfield  {author} {\bibinfo {author} {\bibfnamefont {L.~M.}\ \bibnamefont
  {Sieberer}}\ and\ \bibinfo {author} {\bibfnamefont {W.}~\bibnamefont
  {Lechner}},\ }\bibfield  {title} {\bibinfo {title} {Programmable
  superpositions of ising configurations},\ }\href
  {https://doi.org/10.1103/PhysRevA.97.052329} {\bibfield  {journal} {\bibinfo
  {journal} {Phys. Rev. A}\ }\textbf {\bibinfo {volume} {97}},\ \bibinfo
  {pages} {052329} (\bibinfo {year} {2018})}\BibitemShut {NoStop}%
\bibitem [{\citenamefont {Dlaska}\ \emph {et~al.}(2019)\citenamefont {Dlaska},
  \citenamefont {Sieberer},\ and\ \citenamefont {Lechner}}]{Dlaska2019}%
  \BibitemOpen
  \bibfield  {author} {\bibinfo {author} {\bibfnamefont {C.}~\bibnamefont
  {Dlaska}}, \bibinfo {author} {\bibfnamefont {L.~M.}\ \bibnamefont
  {Sieberer}},\ and\ \bibinfo {author} {\bibfnamefont {W.}~\bibnamefont
  {Lechner}},\ }\bibfield  {title} {\bibinfo {title} {Designing ground states
  of hopfield networks for quantum state preparation},\ }\href
  {https://doi.org/10.1103/PhysRevA.99.032342} {\bibfield  {journal} {\bibinfo
  {journal} {Phys. Rev. A}\ }\textbf {\bibinfo {volume} {99}},\ \bibinfo
  {pages} {032342} (\bibinfo {year} {2019})}\BibitemShut {NoStop}%
\bibitem [{\citenamefont {Farhi}\ \emph {et~al.}(2014)\citenamefont {Farhi},
  \citenamefont {Goldstone},\ and\ \citenamefont {Gutmann}}]{Farhi2014}%
  \BibitemOpen
  \bibfield  {author} {\bibinfo {author} {\bibfnamefont {E.}~\bibnamefont
  {Farhi}}, \bibinfo {author} {\bibfnamefont {J.}~\bibnamefont {Goldstone}},\
  and\ \bibinfo {author} {\bibfnamefont {S.}~\bibnamefont {Gutmann}},\
  }\href@noop {} {\bibinfo {title} {A quantum approximate optimization
  algorithm}} (\bibinfo {year} {2014}),\ \Eprint
  {https://arxiv.org/abs/1411.4028} {arXiv:1411.4028 [quant-ph]} \BibitemShut
  {NoStop}%
\bibitem [{\citenamefont {Ender}\ \emph {et~al.}(2022)\citenamefont {Ender},
  \citenamefont {Messinger}, \citenamefont {Fellner}, \citenamefont {Dlaska},\
  and\ \citenamefont {Lechner}}]{Ender2022}%
  \BibitemOpen
  \bibfield  {author} {\bibinfo {author} {\bibfnamefont {K.}~\bibnamefont
  {Ender}}, \bibinfo {author} {\bibfnamefont {A.}~\bibnamefont {Messinger}},
  \bibinfo {author} {\bibfnamefont {M.}~\bibnamefont {Fellner}}, \bibinfo
  {author} {\bibfnamefont {C.}~\bibnamefont {Dlaska}},\ and\ \bibinfo {author}
  {\bibfnamefont {W.}~\bibnamefont {Lechner}},\ }\bibfield  {title} {\bibinfo
  {title} {Modular parity quantum approximate optimization},\ }\href
  {https://doi.org/10.1103/PRXQuantum.3.030304} {\bibfield  {journal} {\bibinfo
   {journal} {PRX Quantum}\ }\textbf {\bibinfo {volume} {3}},\ \bibinfo {pages}
  {030304} (\bibinfo {year} {2022})}\BibitemShut {NoStop}%
\bibitem [{\citenamefont {Liu}\ \emph {et~al.}(2022)\citenamefont {Liu},
  \citenamefont {Gao}, \citenamefont {Cain}, \citenamefont {Lukin},\ and\
  \citenamefont {Wang}}]{Jin-Guo2022}%
  \BibitemOpen
  \bibfield  {author} {\bibinfo {author} {\bibfnamefont {J.-G.}\ \bibnamefont
  {Liu}}, \bibinfo {author} {\bibfnamefont {X.}~\bibnamefont {Gao}}, \bibinfo
  {author} {\bibfnamefont {M.}~\bibnamefont {Cain}}, \bibinfo {author}
  {\bibfnamefont {M.~D.}\ \bibnamefont {Lukin}},\ and\ \bibinfo {author}
  {\bibfnamefont {S.-T.}\ \bibnamefont {Wang}},\ }\href@noop {} {\bibinfo
  {title} {Computing solution space properties of combinatorial optimization
  problems via generic tensor networks}} (\bibinfo {year} {2022}),\ \Eprint
  {https://arxiv.org/abs/2205.03718} {arXiv:2205.03718} \BibitemShut {NoStop}%
\bibitem [{\citenamefont {Nguyen}\ \emph {et~al.}(2022)\citenamefont {Nguyen},
  \citenamefont {Liu}, \citenamefont {Wurtz}, \citenamefont {Lukin},
  \citenamefont {Wang},\ and\ \citenamefont {Pichler}}]{Nguyen2022}%
  \BibitemOpen
  \bibfield  {author} {\bibinfo {author} {\bibfnamefont {M.-T.}\ \bibnamefont
  {Nguyen}}, \bibinfo {author} {\bibfnamefont {J.-G.}\ \bibnamefont {Liu}},
  \bibinfo {author} {\bibfnamefont {J.}~\bibnamefont {Wurtz}}, \bibinfo
  {author} {\bibfnamefont {M.~D.}\ \bibnamefont {Lukin}}, \bibinfo {author}
  {\bibfnamefont {S.-T.}\ \bibnamefont {Wang}},\ and\ \bibinfo {author}
  {\bibfnamefont {H.}~\bibnamefont {Pichler}},\ }\href@noop {} {\bibinfo
  {title} {Quantum optimization with arbitrary connectivity using {R}ydberg
  atom arrays}} (\bibinfo {year} {2022}),\ \Eprint
  {https://arxiv.org/abs/2209.03965} {arXiv:2209.03965} \BibitemShut {NoStop}%
\bibitem [{\citenamefont {Lanthaler}\ and\ \citenamefont
  {Lechner}(2021)}]{Lanthaler2021}%
  \BibitemOpen
  \bibfield  {author} {\bibinfo {author} {\bibfnamefont {M.}~\bibnamefont
  {Lanthaler}}\ and\ \bibinfo {author} {\bibfnamefont {W.}~\bibnamefont
  {Lechner}},\ }\bibfield  {title} {\bibinfo {title} {Minimal constraints in
  the parity formulation of optimization problems},\ }\href
  {https://doi.org/10.1088/1367-2630/ac1897} {\bibfield  {journal} {\bibinfo
  {journal} {New Journal of Physics}\ }\textbf {\bibinfo {volume} {23}},\
  \bibinfo {pages} {083039} (\bibinfo {year} {2021})}\BibitemShut {NoStop}%
\bibitem [{\citenamefont {Parisi}(1980)}]{Parisi_1980}%
  \BibitemOpen
  \bibfield  {author} {\bibinfo {author} {\bibfnamefont {G.}~\bibnamefont
  {Parisi}},\ }\bibfield  {title} {\bibinfo {title} {A sequence of approximated
  solutions to the s-k model for spin glasses},\ }\href
  {https://doi.org/10.1088/0305-4470/13/4/009} {\bibfield  {journal} {\bibinfo
  {journal} {Journal of Physics A: Mathematical and General}\ }\textbf
  {\bibinfo {volume} {13}},\ \bibinfo {pages} {L115} (\bibinfo {year}
  {1980})}\BibitemShut {NoStop}%
\bibitem [{\citenamefont {Barredo}\ \emph {et~al.}(2018)\citenamefont
  {Barredo}, \citenamefont {Lienhard}, \citenamefont {de~L{\'e}s{\'e}leuc},
  \citenamefont {Lahaye},\ and\ \citenamefont {Browaeys}}]{Barredo2018}%
  \BibitemOpen
  \bibfield  {author} {\bibinfo {author} {\bibfnamefont {D.}~\bibnamefont
  {Barredo}}, \bibinfo {author} {\bibfnamefont {V.}~\bibnamefont {Lienhard}},
  \bibinfo {author} {\bibfnamefont {S.}~\bibnamefont {de~L{\'e}s{\'e}leuc}},
  \bibinfo {author} {\bibfnamefont {T.}~\bibnamefont {Lahaye}},\ and\ \bibinfo
  {author} {\bibfnamefont {A.}~\bibnamefont {Browaeys}},\ }\bibfield  {title}
  {\bibinfo {title} {{Synthetic three-dimensional atomic structures assembled
  atom by atom}},\ }\href {https://doi.org/10.1038/s41586-018-0450-2}
  {\bibfield  {journal} {\bibinfo  {journal} {Nature}\ }\textbf {\bibinfo
  {volume} {561}},\ \bibinfo {pages} {79} (\bibinfo {year} {2018})}\BibitemShut
  {NoStop}%
\end{thebibliography}
%

\end{document}